\documentclass[twocolumn]{article}

\usepackage[utf8]{inputenc}

\usepackage{booktabs} %
\usepackage{makecell}
\usepackage{hyperref}
\usepackage{adjustbox}
\usepackage{multirow}
\usepackage{graphicx}
\usepackage{subcaption}
\usepackage{array}
\usepackage{csquotes}
\usepackage{balance}
\usepackage{enumitem}
\usepackage{xcolor}
\usepackage{tabularx}
\usepackage[acronym,shortcuts]{glossaries}
\usepackage[normalem]{ulem}
\usepackage{listings}
\usepackage[noblocks]{authblk}
\usepackage{amssymb}
\usepackage{xcolor}  %
\definecolor{OwnAzure}{HTML}{336699}
\definecolor{OwnCerulean}{HTML}{CAE2FE}
\definecolor{OwnOliveGreen}{HTML}{556B2F}
\definecolor{KamPurple}{HTML}{907C97}
\usepackage[colorinlistoftodos,prependcaption,textsize=tiny]{todonotes}%
\usepackage{xargs}                      %
\newcommandx{\todoai}[2][1=]{\todo[inline,linecolor=OwnAzure,backgroundcolor=OwnCerulean,bordercolor=OwnAzure,#1]{#2}}

\newacronym{pdf}{PDF}{probability density function}
\newacronym{cdf}{CDF}{cumulative distribution function}
\newacronym{cdfs}{CDFs}{cumulative distribution functions}
\newacronym{cp}{CP}{critical path}
\newacronym{wta}{WTA}{Workflow Trace Archive}
\newacronym{nfr}{NFR}{non-functional requirement}
\newacronym{qos}{QoS}{Quality of Service}
\newacronym{dag}{DAG}{directed acyclic graph}

\hypersetup{draft}

\newcommand{\vcutS}{\vspace*{-0.15cm}}
\newcommand{\vcutM}{\vspace*{-0.25cm}}
\newcommand{\vcutL}{\vspace*{-0.5cm}}
\newcommand{\vcutXL}{\vspace*{-0.75cm}}

\newcommand{\resquestion}[1]{{RQ-#1}}
\newcommand{\wtareq}[1]{{\bf R-#1}}

\newcounter{mainfindingid}
\newcommand{\maindfinding}[2]{\refstepcounter{mainfindingid}\label{#1}\item[O-\arabic{mainfindingid}:] #2}

\newcounter{mainchallengeid}
\newcommand{\mainchallenge}[2]{\refstepcounter{mainchallengeid}\label{#1}{\bf C-\arabic{mainchallengeid}. #2}}

\begin{document}
	
\title{The Workflow Trace Archive: Open-Access Data from\\ Public and Private Computing Infrastructures -- Technical Report}

\date{}

\makeatletter \renewcommand\AB@affilsepx{, \protect\Affilfont} \makeatother

\author[1]{Laurens Versluis}
\author[2]{Roland Mathá}
\author[3]{Sacheendra Talluri}
\author[1]{Tim Hegeman}
\author[4]{Radu Prodan}
\author[5]{Ewa Deelman}
\author[1]{Alexandru Iosup}

\affil[1]{Vrije Universiteit Amsterdam}
\affil[2]{University of Innsbruck}
\affil[3]{Delft University of Technology}
\affil[4]{University of Klagenfurt}
\affil[5]{University of Southern California}
\setcounter{Maxaffil}{0}

\maketitle

\begin{abstract}

Realistic, relevant, and reproducible experiments often need input traces collected from real-world environments. We focus in this work on traces of workflows---common in datacenters, clouds, and HPC infrastructures. We show that the state-of-the-art in using workflow-traces raises important issues: (1) the use of realistic traces is infrequent, and (2) the use of realistic, {\it open-access} traces even more so. Alleviating these issues, we introduce the Workflow Trace Archive (WTA), an open-access archive of workflow traces from diverse computing infrastructures and tooling to parse, validate, and analyze traces. The WTA includes ${>}48$ million workflows captured from ${>}10$ computing infrastructures, representing a broad diversity of trace domains and characteristics. To emphasize the importance of trace diversity, we characterize the WTA contents and analyze in simulation the impact of trace diversity on experiment results. 
Our results indicate significant differences in characteristics, properties, and workflow structures between workload sources, domains, and fields.

\end{abstract}

\vspace*{-0.2cm}
\section{Introduction}
\label{sct:introduction}

{\it Workflows}, that is, applications that organize tasks with data and computational inter-dependencies, are already a significant part of private datacenter and public cloud infrastructures~\cite{ilyushkin2017experimental,wu2015workflow}. 
This trend is likely to intensify~\cite{isom2012your,deelman2018future}, as organizations and companies transition from basic to increasingly more sophisticated cloud-based services. For example, 96\% of companies responding to RightScale's 2018 survey are using the cloud~\cite{weins2018cloud}, up from 86\% in 2012~\cite{86percent2017Meghan}; the average organization combines services across five {\it public and private} clouds.

\begin{figure}
	\centering
	\includegraphics[width=0.23\columnwidth]{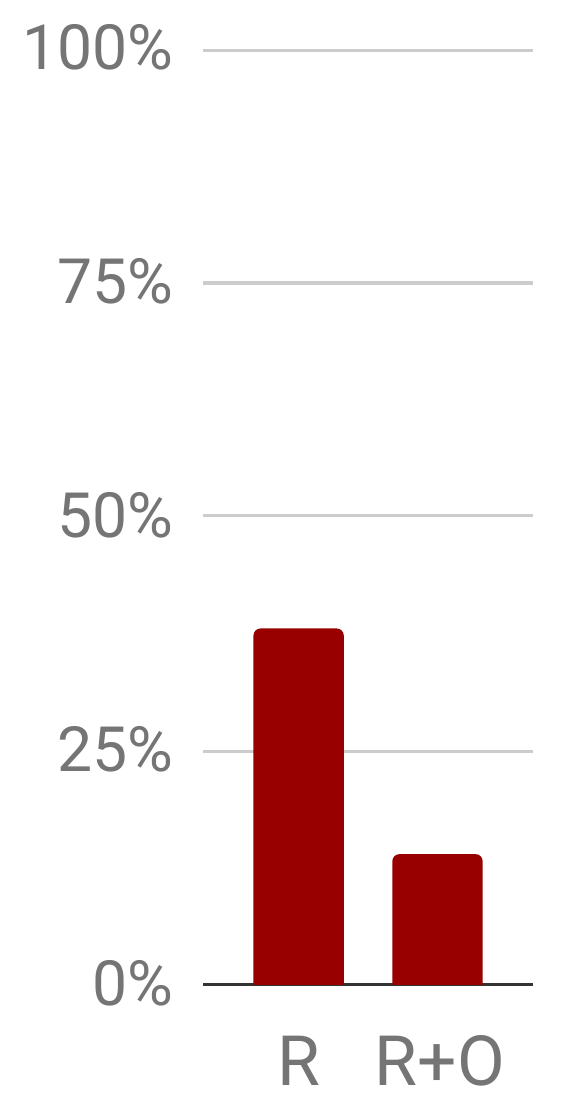}
	\includegraphics[width=0.75\columnwidth]{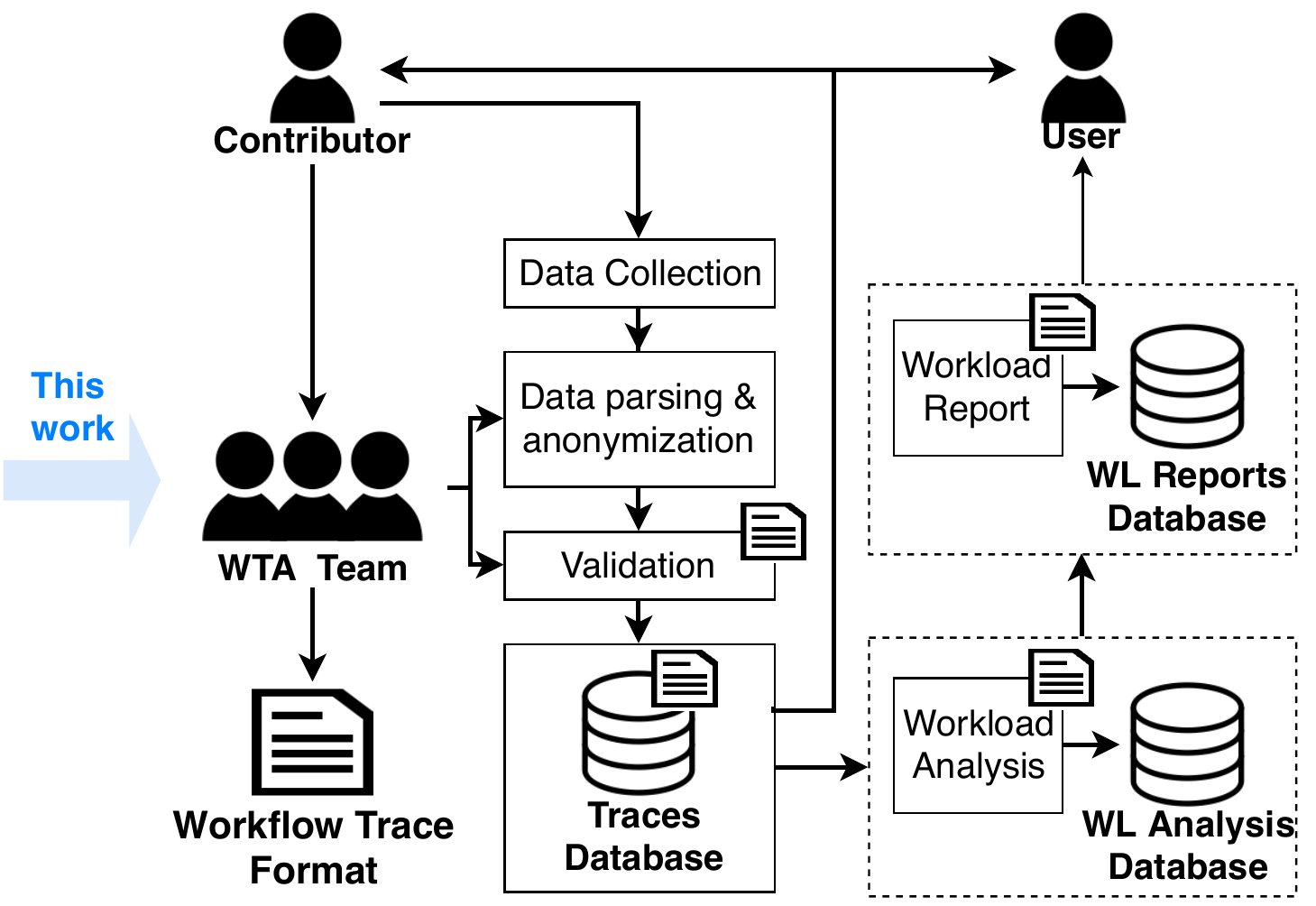}
	\caption{A visual map to this work: ({\it left}) The problem: infrequent use of {\tt R}ealistic ($\approx40\%$) and {\tt O}pen-source ($\approx15\%$) workflow-traces in representative articles, which	can affect the relevance and reproducibility of experiments for the entire community. ({\it right}) Toward an answer: the \gls{wta} stakeholders, process, and tools provide the community with open-source traces of relevant workflows running in public and private computing infrastructures.}
	\label{fig:wta:firstpage}\label{fig:wta:problem}\label{fig:wta:solution}\label{fig:wta-stakeholders}
	\vcutL
\end{figure}

To maintain, tune, and develop the {\it computing infrastructures} for running workflows at the massive scale and with the diversity suggested by these trends, the systems community requires adequate capabilities for testing and experimentation. %
Although the community is aware that workload traces enable a broad class of realistic, relevant, and reproducible experiments, currently such traces are infrequently used, as we summarize in Figure~\ref{fig:wta:firstpage}~(left) and quantify in Section~\ref{sct: survey_trace_usage}.
Toward addressing this problem, we focus on improving trace availability and understanding by proposing a new, free and open-access {\it \gls{wta}}, as detailed in Figure~\ref{fig:wta:firstpage}~(right) and in the remainder of this work. %

This work aligns with our vision of massivizing computer systems (MCS), of ``a world where individuals and human-centered organizations are augmented by an automated, sustainable layer of technology, at the core of which are modern (distributed) computer systems''~\cite{DBLP:conf/icdcs/IosupUVAEHTBT18}.
MCS aims to ``understand and eventually control'' these systems, and posits that, to avoid a reproducibility crisis, the community must ``understand and create together a science, practice, and culture of computer ecosystems'' (principle P8 of MCS) and  must become ``aware of the evolution and emergent behavior of computer ecosystems'' (P9).
MCS also proposes a series of methodological challenges related to these principles, from which this work helps alleviate mainly challenge C19, related to ``understanding and explaining new modes of use, including new, realistic, accurate, yet tractable models of workloads and environments'', but also the other challenges where meaningful workloads currently play a role, such as C15, related to experimentation; C16, related to benchmarking; C17, related to testing, validation, and verification; and C18, related to building a science capable to address modern distributed systems.

The need for workflow traces is stringent~\cite{DBLP:conf/icdcs/IosupUVAEHTBT18,deelman2018future}. 
MCS raises the challenge of understanding all types of workloads that appear in public and private computing infrastructures.
Not only the sheer volume of workloads has increased significantly over time~\cite{wu2015workflow}, but also the users of datacenters and cloud operations are expecting increasingly better \gls{qos} from the workflow-management systems, including elasticity, reliability, and low-cost, under strong assumptions of validation~\cite{DBLP:conf/icdcs/IosupUVAEHTBT18,deelman2018future} and reproducibility~\cite{isom2012your, gil2007examining}. Developing workflow management systems to meet these requirements requires considerable scientific and technical advances and, correspondingly, comprehensive trace-based experimentation and testing, %
both {\it in vivo} and {\it in silico}~\cite{dudley2010silico}.
Testing such systems, especially at cluster and datacenter scale, often cannot be done in vivo, due to downtime or the operational costs required.
Instead, workflow traces can be replayed in silico, allowing multiple setups to run in parallel, testing individual components, etc. without the downtime nor costs.
Although {\it realistic workflow traces} are key for testing, tuning, validating, and inspiring system designs, they are currently still scarce~\cite{cohen2011search}.
Prior work, such as WorkflowHub~\cite{da2014community}, has introduced numerous workflow traces, yet only from the science domain. %
As Figure~\ref{fig:wta:firstpage}~(left) indicates, and Section~\ref{sct: survey_trace_usage} quantifies and explains, less than 40\% of relevant articles focusing on workflow systems conduct experiments with {\it realistic} traces, and less than 15\% conduct experiments with realistic and {\it open-source} traces.

The current scarcity of traces forces researchers to either use synthetically generated workloads or to use one of the few available traces.
Synthetic traces may reduce the representatives and quality of experiments, if they do not match relevant real-world settings.
Using realistic traces that correspond to a narrow application-domain may result in overfitting; Amvrosiadis et al.~\cite{amvrosiadis2018diversity} demonstrate this for cluster-based infrastructures.
Additionally, a lack of realistic traces may lead to limited or even wrong understanding of workflow characteristics, their performance, and their usage, which hampers the reuse of the  systems tested with such (workloads of) workflows~\cite{ramakrishnan2010multi}.
This gives rise to the research question \textit{\resquestion{1}: How diverse are the workflow traces currently used by the systems community?} %

We identify the need to share workflow traces collected from relevant environments running relevant workloads under relevant constraints.
Effective sharing requires unified trace formats, and also support for emerging and new features. 
For example, since the introduction of commercial clouds, clients have increasingly started to ask for better \gls{qos}, and in particular have started to increasingly express \glspl{nfr} such as availability, privacy, and security demands in traces~\cite{deelman2018future, cardoso2002workflow}. 
This leads us to research question \textit{\resquestion{2}: How to support sharing workflow traces in a common, unified format? How to support in it arbitrary \glspl{nfr}?} %

Persuading both academia and industry to release data is vital to address the problems stated prior. We tackle this issue with two main approaches. First, by offering tools to obscure sensitive information, while still retaining significant detail in shared traces. Second, by encouraging the same organization to share the data across its possibly multiple workflow management systems ({\it sources}), and by explicitly aiming to collect data across diverse application {\it domains} and {\it fields}. 
The availability of diverse data and tools stimulate the benefits of making available such traces, while simultaneously reducing the concerns of competitive disadvantage or of an (accidental) disclosure of sensitive information.
The community is already helping with both approaches, by increasingly focusing on the problem of reproducibility.
For example, ACM introduced artifact review and badges to stimulate open-access software and data artifacts for reproducibility and verification purposes~\cite{result2017artifact}.
We add to this community-effort ours, which is scientific in nature: \textit{\resquestion{3}: What is the impact of the source and domain of a trace on the characteristics of workflows?}

Addressing research questions~1--3, our contribution is four-fold:\vcutS
\begin{enumerate}[leftmargin=*,align=left]
	\item To answer \resquestion{1}, we conduct the first comprehensive survey of how the systems community uses workflow traces (Section~\ref{sct: survey_trace_usage}).
	We collect, select, and label articles from top conferences and journals covering workflow management.
	We analyze the types of traces used in the community, and the domains and fields covered in published studies.

	\item To answer \resquestion{2}, we design the \gls{wta} for open-access to traces of {\it workloads} of workflows (Section~\ref{sct:workflow-trace-archive}). 
	We identify a comprehensive set of requirements for a workflow trace archive.
	A key conceptual contribution of the \gls{wta} is the design of a unified trace format for sharing workflows, the first to generalize \glspl{nfr} support at both workflow- and task-levels.
	The \gls{wta} currently archives a diverse set of (1) real workflow traces collected from real-world environments, (2) realistic workflow traces used in peer-reviewed publications, and (3) workflow traces collected from simulated and emulated environments commonly used by the systems community. %
	\gls{wta} also introduces tools to detail and compare its traces. %
	
	\item To address \resquestion{3}, we compare key workload characteristics across traces, domains, and sources (Section~\ref{sct:workflow-characterization}).
	Our effort is the first to characterize the new trace from Alibaba, and the first to investigate the critical path task length, level of parallelism, and burstiness using the Hurst exponent on workloads of workflows.
	Overall, the archive comprises 95 traces, featuring more than 48 million workflows containing over 2 billion CPU core hours.
	
	\item To validate our answers to RQs~1--3, we analyze various threats~(Section~\ref{sct:validation-and-threats}). %
	We conduct a trace-based simulation study and qualitative analysis.
	Our results for the former 
	indicate systems should be tested with different traces to validate claims about the generality of the proposed approach.
	
\end{enumerate}
\section{A Survey of Workflow Trace Usage}
\label{sct: survey_trace_usage}

\begin{table*}[]
	\setlength{\tabcolsep}{4pt}
	\centering
	\caption{Workflow trace usage in venues having at least one paper returned in the initial query. The venues with $>5$ hits have their individual column. The column ``Other'' shows combined results for conferences with $\leq 5$ hits: ATC, CLOUD, CLUSTER, e-Science, Euro-Par, GRID, HPDC, JSSPP, IC2E, ICDCS, ICPE, IPDPS, NSDI, OSDI, {\tt SC}, SIGMETRICS, WORKS.
	Percentages 
	are computed from the total in the corresponding column, e.g., 13 out of 37 for the cell corresponding to row {\tt R} and column {\tt FGCS}.} \label{tbl:workflow-trace-usage-venues}
	\vcutM
	\adjustbox{max width=\linewidth}{
	\begin{tabular}{clrrrrr} \toprule
		& Acronym & \multicolumn{1}{l}{Total} & \multicolumn{1}{l}{FGCS} & \multicolumn{1}{l}{CCGrid} & \multicolumn{1}{l}{TPDS} & \multicolumn{1}{l}{Other}\\ \midrule
		{\tt T} & Articles using traces & \multicolumn{1}{l}{104} & \multicolumn{1}{l}{37} & \multicolumn{1}{l}{17}  & \multicolumn{1}{l}{17}  & \multicolumn{1}{l}{33} \\
		\midrule
		{\tt R} & Articles using \emph{realistic} traces & 40 (38\%) & 13 (35\%) & 8 (47\%) & 6 (35\%) & 13 (39\%)\\
		{\tt R+O} & Articles using traces that are both {\it realistic} and {\it open-access} & 14 (13\%) & 6 (16\%) & 2 (12\%) & 3 (18\%) & 3 (9\%)\\
        \bottomrule
	\end{tabular}
    }%
	\vcutM
\end{table*}

\begin{figure}[t]
	\centering
	\includegraphics[width=0.9\columnwidth]{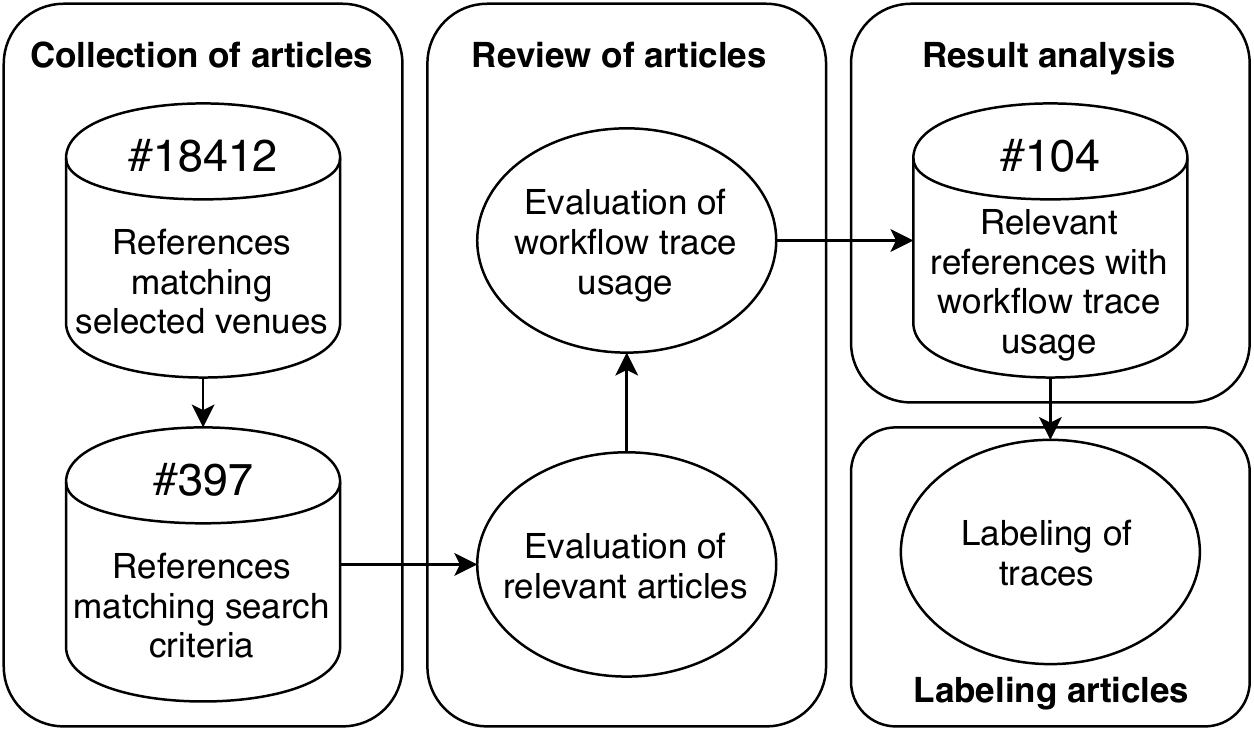}
	\caption{The article selection process. 
	Subsequent stages decrease the amount of articles: from a corpus of 18,412 articles, down to 104 relevant references.} \label{fig:paper-selection-process}
	\vcutM
\end{figure}

To assess the current usage of workflow traces in the systems community and the need for a workflow archive, we systematically survey a large body of work published in top conferences and journals, and investigate articles that perform experiments using workflow traces, either through simulation or using a real-world setup.
The process and outcome of this survey answer \resquestion{1}.
\vcutS

\subsection{Article Selection and Labeling}\label{sec:survey:process}

{\bf Selection:} Figure~\ref{fig:paper-selection-process} displays our systematic approach to select articles relevant to this survey, based on~\cite{kitchenham2009systematic}.
First, we collect data from DBLP~\cite{ley2002dblp} and Semantic Scholar~\cite{ammar:18}. We filter them by venue, retaining only articles from the 10 key conferences and journals in distributed systems listed in the caption of Table~\ref{tbl:workflow-trace-usage-venues}, including {\tt SC}. This yields 18,412 articles. 
Next, we automatically select all articles from the last decade (2009--2018) containing the word \enquote{workflow} in either title or abstract, yielding 397 articles.
This step provides articles that focus on all aspects of workflows, e.g. scheduling, analysis, and design.
Finally, to obtain insights into workflow traces usage, we manually check the 397 articles. 
Overall, this systematic process yields 104~articles using workflow traces. %
To highlight the relevance of papers, we use Google Scholar to obtain citation counts. In total the 104 papers have been cited 3,965 times.

{\bf Labeling:} 
We label for each of the 104 articles the type of trace usage. 
For articles  explicitly describing their use, we use the labels {\it realistic} for traces derived from real-world workflow executions. We label the others as {\it synthetic}. We further label traces as {\it open-access} (or open-source) if they are available online and to a broad audience, and {\it closed-access} (or closed-sources) otherwise. In our analysis, we include among the open-access traces only those that are also realistic.

We also label articles by {\it domain} and {\it field}. We identify in articles explicit use of traces from the domains ``scientific'', ``engineering'', ``multimedia'', ``governmental'', and ``industry'', and from fields such as ``bioinformatics'', ``astronomy'', ``physics'', etc.
We further label a trace with {\it uncategorized} when its origin remains unexplained. %
\vcutS

\subsection{Types of Traces Used in the Community}\label{sec:survey:type}

We analyze here the types of traces used by the community, with the following {Observations (Os)}:
\begin{description}
	\maindfinding{mf:use-realistic}{Less than 40\% of articles use realistic traces.} %
	\maindfinding{mf:use-open-traces}{Only one-seventh of all articles use open-access traces.}
\end{description}

Table~\ref{tbl:workflow-trace-usage-venues} presents the types of traces used in the community, focusing on realistic ({\tt R}) and open-access ({\tt R+O}) traces. 
The community uses traces for experiments across both conference and journal articles, across various levels of (high) quality.
In contrast to this positive finding, the results indicate that, from the total number of articles using traces at all, the fraction of articles using realistic and even open-access traces is relatively small.
Across all venues, only 38\% of the articles use at least one realistic trace, and only 14\% of the articles use at least one open-access trace.

These findings match the perceived difficulty in reproducing studies in the field~\cite{gil2007examining, dudley2010silico}, and may hint why so few of these seemingly successful designs are adopted for use in practice~\cite{schwiegelshohn2014design}.
\vcutS

\begin{figure}
	\centering
	\includegraphics[width=\columnwidth]{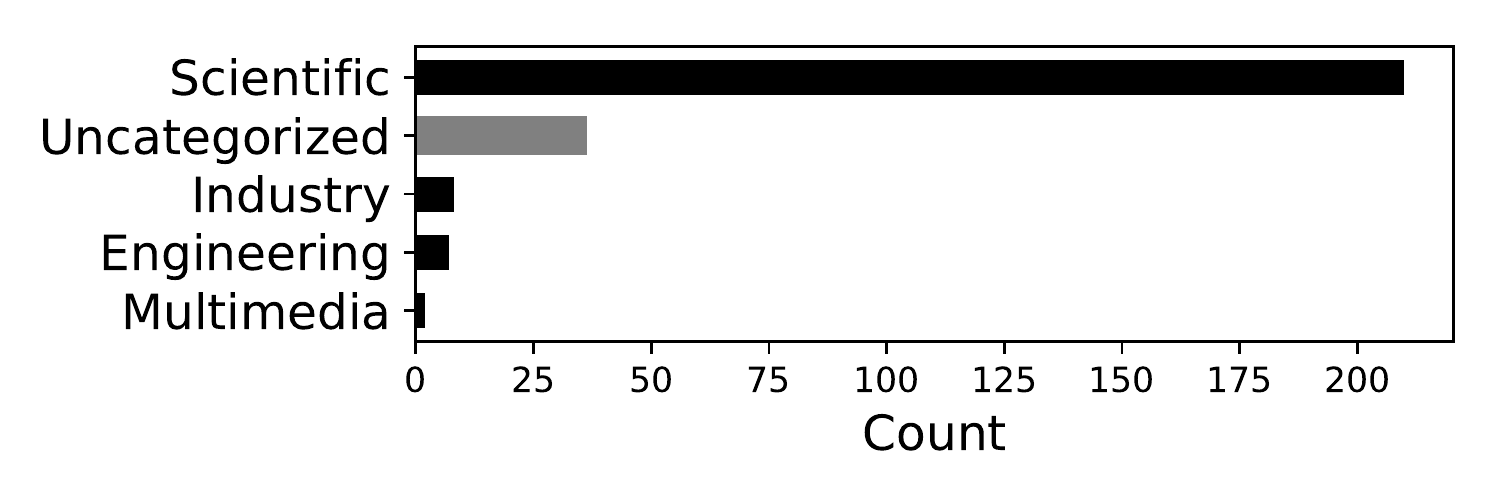}
	\vcutM
	\includegraphics[width=\columnwidth]{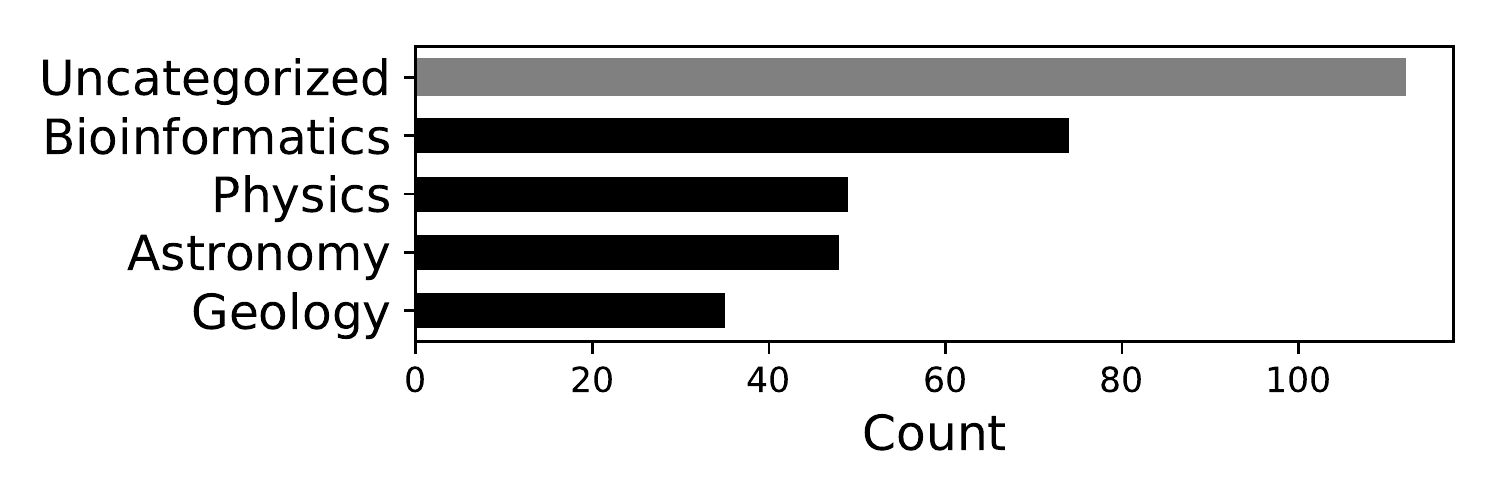}
	\vcutL
	\caption{({\it top}) Top-5 (out of 6) domains and ({\it bottom}) Top-5 (out of 28) fields from which the community sources workflows. (``Uncategorized'' for unclear domain or field.)} \label{fig:trace-domain-count}\label{fig:fields-count}
	\vcutL
\end{figure}

\pagebreak
\subsection{Workflow Domains and Fields}\label{sec:survey:domain}\label{sec:survey:field}\label{sec:survey:diversity}

We analyze the domains and fields from which the community sources workflows, with as main observations:
\begin{description}
    \maindfinding{mf:domain-fields}{The community sources workflows from 5+ domains and 25+ fields.}
	\maindfinding{mf:use-scientific-domain}{Traces containing scientific workflows are used significantly more (20x) than workflows from other domains, e.g., industry and engineering, in the surveyed articles.}
    \maindfinding{mf:bioinformatics-common}{Bioinformatics workflows are the most commonly used, but three other fields exhibit usage within a factor of 3.}
    \maindfinding{mf:uncategorized-often}{Many traces have uncategorized domain and/or field.}
\end{description} 

Overall, we find that the community uses diverse workflows, sourced from 6 domains and 28 fields. We further investigate the distribution of use, per domain and per field.
Figure~\ref{fig:trace-domain-count}~(top) shows
that the scientific domain is over-represented in the literature in the top-five trace domains encountered,
due to the large number of available open-access traces and from their conventional use in prior work.
In particular, a large portion of the articles use workflow traces from the Pegasus project, which covers the scientific domain. The number of traces in this domain exceeds 200, which is larger than the number of articles in the study as each article uses multiple traces.
In contrast, the next-largest domains 
are industry and engineering, each with less than 10 traces representing less than one-twentieth of the scientific domain. 

We remark the positive diversity of the workflow domains, considering that the entire community is tempered by the extreme focus on scientific workflows. This confirms the bias demonstrated by Amvrosiadis et al.~\cite{amvrosiadis2017bigger} with the popular Google-cluster traces. A similar situation appears for fields, but more tempered, as Figure~\ref{fig:fields-count}~(bottom) indicates. %

Overall, the results reveal that the community has a strong bias for one domain (scientific) and favors scientific fields (especially bioinformatics). 
We conjecture the large amount of open-access data from these fields facilitates this bias.
To overcome it, and to further reduce the large fraction of uncategorized traces evident in both plots of Figure~\ref{fig:fields-count}, we posit the community should require that open-access and {\it diverse} traces be used in articles claiming the generality of their techniques. %
\vcutS

\begin{figure}[t]
	\begin{subfigure}{\columnwidth}
		\centering
		\includegraphics[width=0.4\textwidth]{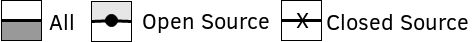}
	\end{subfigure}
	\begin{subfigure}{\columnwidth}
		\centering
		\includegraphics[width=\textwidth]{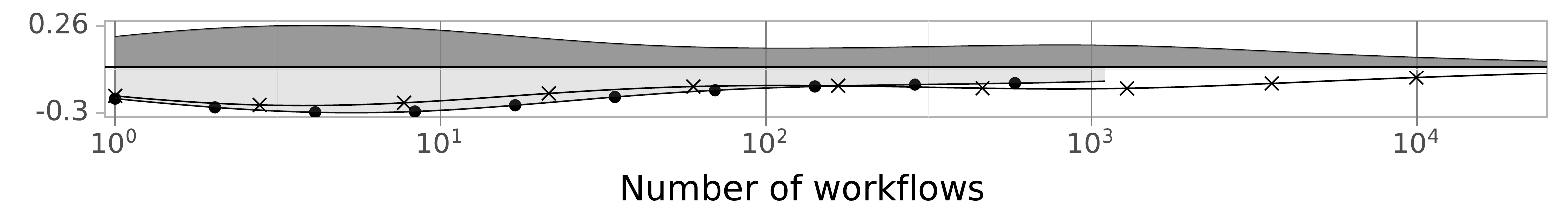}
		\caption{Number of workflows per article.} \label{fig:violinplot-num-wfs-per-experiment}
	\end{subfigure}
	\begin{subfigure}{\columnwidth}
		\centering
		\includegraphics[width=\textwidth]{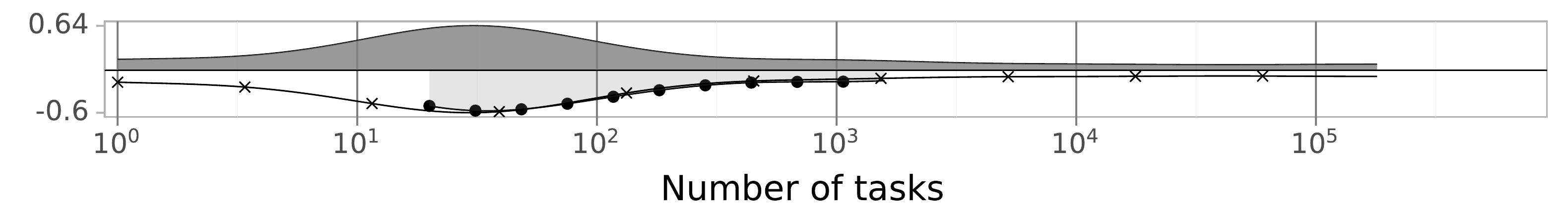}
		\caption{Smallest workflow size per article.} \label{fig:smallest_workflow_distribution}
	\end{subfigure}
	\begin{subfigure}{\columnwidth}
		\centering
		\includegraphics[width=\textwidth]{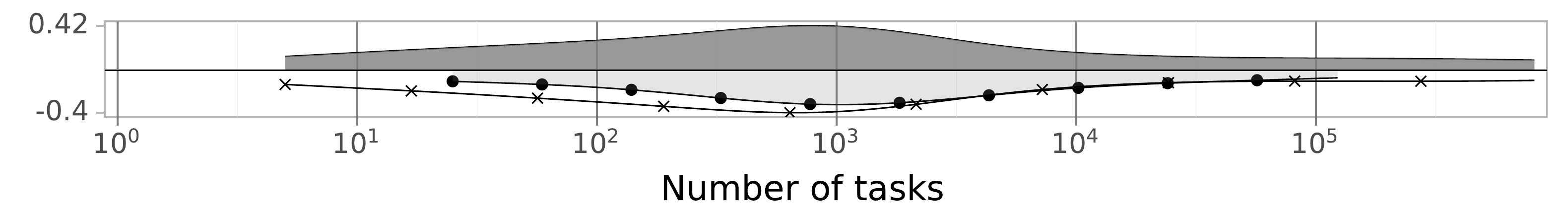}
		\caption{Largest workflow size per article.} \label{fig:largest_workflow_distribution}
	\end{subfigure}
	\caption{Statistical characterization of workflows in open- and closed-access traces. 
	Horizontal axes are logarithmic. Symbols are differentiation markers, not data-points.} %
 \label{fig:workflow_distribution_characteristics}
 \vcutM
\end{figure}

\subsection{Workflow Size in \\Open- and Closed-access Traces}\label{sec:survey:open}

We focus on open- and closed-access traces---is there structural or size-related inequality? We find that:
\begin{description}
	\maindfinding{mf:open-access-fewer-workflows}{Open-access traces generally have fewer workflows.}
	\maindfinding{mf:open-access-different-sizes}{Open/closed-access traces cover different workflow sizes.} %
 \end{description}

Figure~\ref{fig:workflow_distribution_characteristics} uses split violin plots to characterize the number and size of workflows present in open- vs. closed-access traces (bottom half-violins), and across all traces (top). Each half-violin depicts the \gls{pdf} of an empirical variable: number of workflows used per article in Figure~\ref{fig:violinplot-num-wfs-per-experiment}, etc.

Figure~\ref{fig:violinplot-num-wfs-per-experiment} shows that most articles use up to 10 workflows for experiments, with an average of 11 workflows per article and some extreme outliers. 
The distributions for open- and closed-access traces are similarly shaped, yet closed-access traces can include over an order of magnitude more workflows.

For the smallest workflow size, Figures~\ref{fig:smallest_workflow_distribution} indicates that about half of articles use workflows with 30 tasks or fewer.
The range of workflow-sizes in open-access traces is much smaller than for  closed-access traces; in particular, articles using closed-access traces also report many workflows with 10 or less tasks.

 For the largest workflow size, half of experiments do not exceed 1,000 tasks, matching the prior characterization of Juve et al.~\cite{juve2013characterizing}. 
 Although closer than for the smallest workflow size, the open- and close-access traces are again exhibiting different ranges for their largest workflow sizes.

\section{The Workflow Trace Archive}
\label{sct:workflow-trace-archive}

In this section we outline the design of the \gls{wta}, the unified trace format used, tools to support consumers with the trace selection according to their use-case, and give a summarized overview of the current contents of the archive. Furthermore, that facilitate the continuous growth of the archive, we provide tools for trace anonymization and a collection of trace parse scripts for different trace sources.

Similarly to how the overall design of experiments is now commonly described in publications in our field, as the setup leading to experimental results, 
we include here an overview of the design process that led to the design presented in this section. 
Outlining the design and the process that led to the design is important for understanding how the final design came to be and how it fits the intended goal~\cite{2019arXiv190205416I}.

The {\it overall design} of the archive and the {\it detailed design} that results in the unified format are done iteratively, using a process that draws from the AtLarge systematic design-process~\cite{2019arXiv190205416I}.
We started by listing initial requirements (see Section~\ref{sct:wta-requirements}) that the \gls{wta} had to fulfill, and {\it co-evolved the requirements with the development of the solution} (the archive). 
For example, we added explicitly the requirement to provide scripts and datasets to aid users in building their own tools, as we discovered how difficult it was to engineer them from scratch (see Section~\ref{ssct:tools-analysis-and-validation}). 
Next, we defined an initial format, centered around a number of unique features, such as NFRs.
We improved this format iteratively, to meet the requirements and/or to pass various thought experiments. 
For the latter, whenever we encountered a new data-format that was not fully covered by our format, we discussed which properties and/or objects should be added to the format (see Section~\ref{ssct:trace-format}).
We assessed the trade-off between format comprehensiveness (what to include?) and brevity (what is too much or too complex?) based on personal experience, on the perceived importance of data-fields in literature, and on their frequency of use in other archives. 
Finally, we designed the analysis tools iteratively, 
including in them initially our own ideas and then aspects highlighted by other archives, literature reports, and perceived shortcomings. %
\vcutS

\subsection{Requirements}\label{sct:wta-requirements}

We identify five key requirements for the structure, content, and operation of a useful archive for workflow traces.

\noindent \textbf{\wtareq{1}: Diverse Traces for Academia, Industry, and Education.}
The main observations in Section~\ref{sec:survey:diversity} give strong evidence that the archive must include a diverse set of traces, corresponding to the many existing workflow domains and fields. Moreover, the archive must include traces that cover a broad spectrum of workflow sizes, structures, and other characteristics, including both general characteristics to many domains and fields, and idiosyncratic characteristics corresponding to only one domain or field. 

Addressing this requirement is important.
For academia, experimenting with representative traces demonstrates the applicability of an approach, increases credibility.
For industry, representative traces are essential to test production-ready systems, and can act as a validation for techniques proposed by academia~\cite{klusavcek2014interactions}.
Furthermore, as systems become more complex, education and training on these topics becomes ever more important.

\noindent \textbf{\wtareq{2}: A Unified Format for Workload of Workflows Traces.} %

To simplify trace exchange, reduce trace integration effort among different systems, prevent double work for other users, and provide dataset independent tools (expressed as \wtareq{3}), the archive must define a unified trace format.
The format must cover a broad set of data about the workloads and about the workflow management systemsm including:
workload metadata; 
workflow-level data including \glspl{nfr} and common metadata;
task-level data including per-task \glspl{nfr} and operational metadata such as task-state changes;
inter-dependencies between tasks and other operational elements such as data transfers; 
system-level information including resource provisioning, allocation, and consumption; etc.

Addressing this requirement is the basis of any data archive. A unified trace format for workflow workloads helps providing long-term provenance information~\cite{7160272} to improve the reproducibility of experimental results, which is an ongoing grand challenge in fields such as bioinformatics.

\noindent \textbf{\wtareq{3}: Detailed Insights into Trace Properties.}

Just archiving data is not enough; every archive must provide insight into its traces at the level of detail required by the broad audience implied by \wtareq{1}, from beginner to expert.
Broad insights, easily provided in standardized reports, include {\it extrinsic properties} such the number of workflows and tasks and the number of users in the trace, while {\it intrinsic properties} include the workflow arrival patterns and the resource consumption per-task.

Detailed insights include expert-level {\it analysis of single traces} at workload-, workflow-, and system-level; and {\it collective analysis} across all traces and across traces with a shared feature (e.g., all traces of a domain or field).
These properties must be accessible through readily available tools (see \wtareq{4}) and, possibly, through interactive online reports.
Additionally, practitioners can correlate information across multiple traces, resulting in better quantitative evidence, intuition about otherwise black-box applications, and understanding that helps avoiding common pitfalls~\cite{DBLP:conf/jsspp/FrachtenbergF05}.

\noindent \textbf{\wtareq{4}: Tools for Trace Access, Parsing, Analysis, Validation.}

The most important tool is the online presence of the archive itself.
The archive must further provide tools to parse traces from different sources to the unified format~(see also~\wtareq{2}), to provide insight into traces~(see also~\wtareq{3}), and to {\it validate} common properties (e.g., the presence of and correctness of properties).
An absence of such tools would lead to users unable to select appropriate traces, validate their properties, and compare them.

The archive should further aid users in building more sophisticated tools.
Newly built tools can then be added to the selection of tools so more parties can make use of them (contributing to \wtareq{5})

\noindent \textbf{\wtareq{5}: Methods for Contribution.}

The archive must reflect the continuous evolution of workflow use in practice, by increasing the coverage of different scenarios.
We make a distinction between two types of contribution: 
(1) additional traces, possibly originating from a new domain or application-field, and 
(2) additional traces, introducing new properties.
To facilitate the former contribution, the archive must provide a method for the upload and (basic) automated verification of traces. %
To facilitate the latter, the format must integrate specific provisions that enable upgrades and long-term maintainability, such as adding a version to each component of the format.

Having methods to contribute new workloads is key to encourage new and existing contributors to submit traces.
In particular, tools to add new domains are of particular importance, to support emerging paradigms with realistic data.

\subsection{Overview of the \gls{wta}}
\label{sct:overview-wta}

We design the \gls{wta} as a process and set of tools helping a diverse set of stakeholders.
We consider three roles for the \gls{wta} community members, outlined in Figure~\ref{fig:wta-stakeholders}.
The \textit{contributor} supplies, as the legal owner or representative, one or more traces to the \gls{wta}.
A workflow trace contains historical task execution data, resource usage, \glspl{nfr}, resource description, inputs and outputs, etc.
To fulfill \wtareq{5}, the \textit{\gls{wta} team} assists the contributor in parsing, anonymizing, and converting the traces into the unified format (Section~\ref{ssct:trace-format}), minimizing the risk of competitive disadvantage, and verifying their integrity.
\gls{wta} fulfills \wtareq{1} as it incrementally expands with contributors of traces from different domains with different properties.

The \textit{user} represents non-expert or expert trace consumers.
Non-expert users often need to rely on generic domain or trace properties, whereas the expert users have detailed knowledge of their system and require fine-grained detailed for selecting the correct trace. In addition, expert users may comment on (missing) properties and may develop new tools, models or other techniques to further compare and rank the traces.
Both user types require assistance in selecting the most suitable trace given a set of criteria (Section~\ref{ssct:mechanisms-trace-selection}) as well as analysis and validation (Section~\ref{ssct:tools-analysis-and-validation}) from the available set of traces (Section~\ref{ssct:current-content}).
To support both user types, the \gls{wta} discloses both high-level and low-level details.
\vcutS

\subsection{Workflow Model}
\label{ssct:workflow-model}
There are numerous types of workflow models used across different communities.
A 2018 study by Versluis et al. find that \glspl{dag} are the most commonly used formalism in computer system conferences~\cite{DBLP:conf/wosp/VersluisEI18}.
Therefore, for the first design of this archive, we adopt \glspl{dag} as the workflow model.

A workflow constructed as a \gls{dag} in which nodes are computational tasks and directed edges depict the computational or data constraints between tasks. 
Entry tasks are tasks with no incoming dependencies and, once submitted to the system, immediately are eligible for execution.
Similarly, end tasks are nodes that have no outgoing edges.
A collection of workflows submitted to the same infrastructure over a certain period of time is considered a \textit{workload}.

Although popular, we specifically do not focus in this work on BPMN and BPEL, Petri nets, hyper graphs, general undirected, or cyclic graphs.
These formalisms either include business and human-in-the-loop elements~\cite{DBLP:journals/jid/AalstWW03} or add additional complexity due to having a large set of control structures such as loops, conditions, etc.~\cite{DBLP:books/sp/07/Slominski07} which we consider out of scope for this work.

\subsection{Unified Trace Format}
\label{ssct:trace-format}

Creating a unified format (\wtareq{2}) requires from the designer a careful balance between limiting the number of recorded fields while supporting a diverse set usage scenarios for all stakeholders in Section~\ref{sct:overview-wta}.
Modern logging and tracing infrastructure can capture thousands of metrics for each machine and workflow-task involved~\cite{DBLP:conf/wosp/XiongPZG13}, from which the designer must select.
We specifically envision support for common system and workflow properties found in the typical scenarios considered in the top venues surveyed in Section~\ref{sct: survey_trace_usage}, such as engineering a workflow engine~\cite{DBLP:conf/hpdc/ZhouHCL15}, characterizing the properties of workloads of workflows~\cite{juve2013characterizing}, and designing and tuning new workflow schedulers~\cite{DBLP:conf/sc/SunJRBZYKKCP15}.

\begin{figure}
	\centering
	\includegraphics[width=\linewidth,height=\textheight,keepaspectratio]{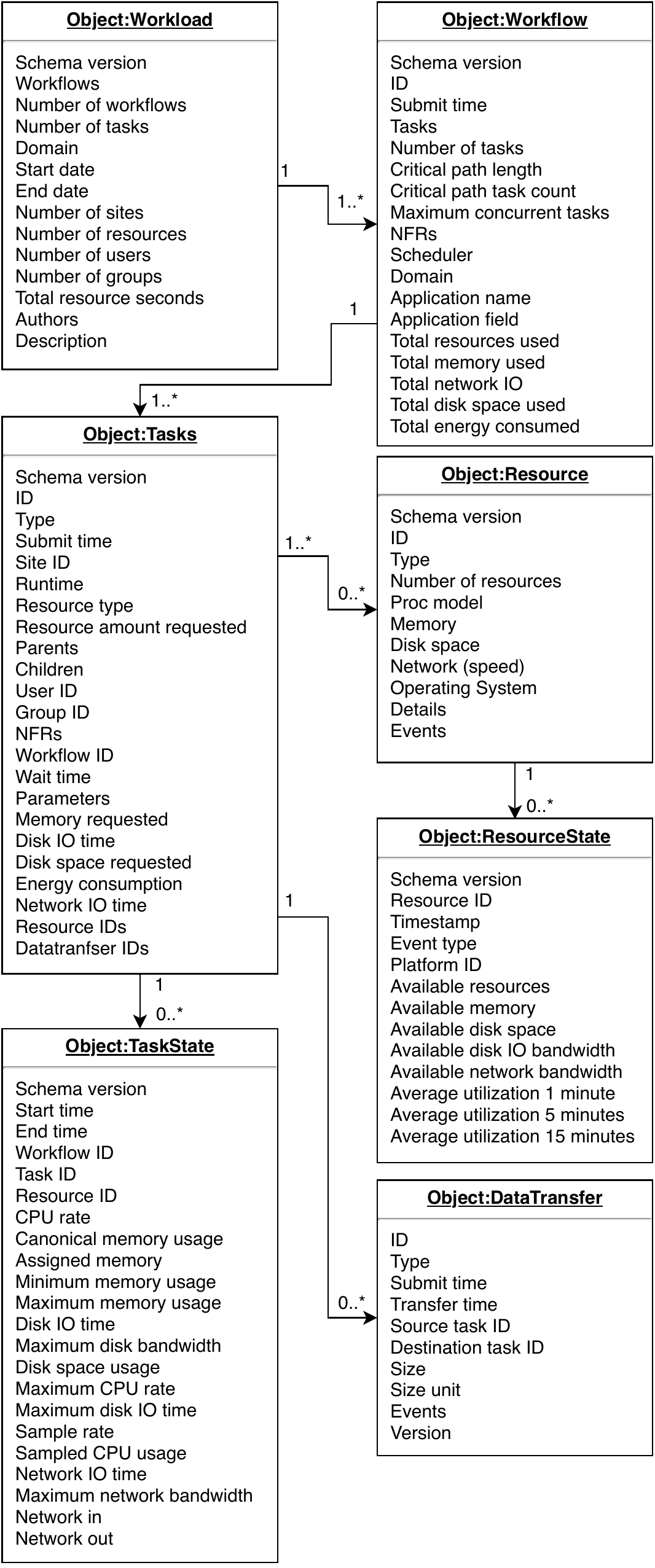}
	\caption{The \gls{wta} trace format.} \label{fig:wta-formalism-asbstract}
	\vcutL
\end{figure}

Our unified format attempts to cover different trace {\it domains}, while preserving valuable information, such as resource consumption and \glspl{nfr}, contributing to fulfilling \wtareq{1 and 3}.
By analyzing the raw data formats, we carefully selected useful properties to include in our unified format, omitting
low-level details, such as cycles per instruction, page cache sizes, etc.

Answering \resquestion{2} and fulfilling \wtareq{2}, our trace format is the first to support arbitrary \glspl{nfr} both at task and workflow levels.
For example, one of the LANL traces (introduced in Table~\ref{tbl:current-content-summary}) contains deadlines per workflow and the Google cluster data features task priorities, both are supported by the \gls{wta} unified format.
Capturing these properties is important to test \gls{qos}-aware schedulers.

As depicted in Figure~\ref{fig:wta-formalism-asbstract}, the \gls{wta} format includes seven objects: \emph{Workload}, \emph{Workflow}, \emph{Task}, \emph{TaskState} \emph{Resource}, \emph{ResourceState}, and \emph{DataTransfer}.
Each of these objects contains a version field, updated whenever the set of properties is altered (\wtareq{5}).

Each trace is a single \emph{workload}, consisting of multiple workflows and their {\it arrival process}.
Workload properties include the number of workflows, tasks, users, resources used (e.g., CPUs or cores), domain and field when available, authors list, start and end date, and overall resource consumption statistics. 
The workload {\it description} described the origin of the workflow and other relevant information, including but not limited to its source, execution environment, non-functional requirements etc.

Each \emph{workflow} in the workload has a unique identifier, an arrival time, and contains a set of tasks and several properties, including scheduler used, number of tasks, critical path length, \glspl{nfr}, and resource consumption. 
Each workflow also has the {\it name} of its {\it domain} of study, when possible.

The \textit{workflow structure}  is a \gls{dag} of tasks with control and data dependencies between tasks.
Each \emph{Task} has a unique identifier and lists its submission and waiting time, runtime and resource requirements, including required  (compute) resources, memory, disk, network, and energy usage.
Additionally, each task provides optional dictionaries for task-specific execution parameters and \glspl{nfr}.
To model dependencies between tasks, the \gls{wta} format maintains for  each task a list of its predecessor (parent) and successor (child) tasks.
Similarly, data dependencies are recorded as a list of data transfers.

\emph{Resource} objects cover various resource types, such as cloud instances, cluster nodes, and IoT devices. 
A resource has a unique identifier and contains several properties, such as resource type (e.g., CPU, GPU, threads), number, processor model, memory, disk space, and operating system. 
An optional dictionary provides further details, such as instance type or Cloud provider.
The \emph{ResourceState} event snapshots periodically the resource state, including availability and utilization.
Analogous to the ResourceState, the \emph{TaskState} records periodically the resource consumption of the task (the Task object records the resource demand).
Different from the ResourceState, the TaskState snapshot contains averages over a period of time.
This is due to the resources generally allocating resources to tasks, whereas the actual consumption of a task may differ over time.

Each \emph{DataTransfer} describes a file transfer from a source to a destination task, which can be a local copy on the same resource or a network transfer from a remote source, etc.
To support bandwidth analysis, a data transfer introduces submission time, transfer time, and data size.
Each data transfer also provides an optional dictionary with detailed event timestamps (e.g., pause, retry).
\vcutS

\subsection{Mechanisms for Trace Selection}
\label{ssct:mechanisms-trace-selection}

\begin{figure}[t]
	\centering
	\includegraphics[width=\columnwidth]{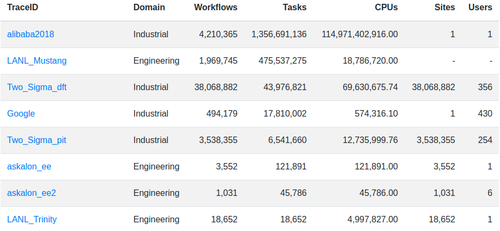}
	\caption{A screenshot of a part of the table containing all traces on the \gls{wta} website, sorted on the tasks column.} \label{fig:wta-website-table}
	\vcutM
\end{figure}

We address \wtareq{3} by assisting archive users in retrieving appropriate traces for their scenarios, using filter and selection mechanisms.
The website is the most important such filter and mechanism, containing
an overview of all traces in a general table, visible in Figure~\ref{fig:wta-website-table}, with the number of workflows, tasks, users, etc. This table is sortable and searchable, allowing website users to interact with the more than 90 traces currently in the \gls{wta} (column ``\#WL'', row ``Total'' in Table~\ref{tbl:current-content-summary}).

We provide, online and as separate tools, a detailed report for each trace. 
Each report includes automatically generated statistics,
such as the number of workflows and tasks, then resource properties such as compute, memory, and IO,
and job and task arrival times and runtime distributions (see Section~\ref{sct:workflow-characterization}).
The metrics featured in the report are reported as important by prior studies~\cite{simmhan2006framework, chen2015using} and enable developers to select traces appropriate for their intended use-case.
\vcutS

\subsection{Tools for Analysis and Validation}
\label{ssct:tools-analysis-and-validation}

\begin{table}[t]
	\centering
	\caption{Trace anonymization methods used in WTA tools.}
	\adjustbox{max width=\columnwidth}{
	\begin{tabular}{ll}
		\toprule
		Obfuscation method & Description \\ 
		\midrule
		IP & Encodes IPv4 addresses\\
		Mail and host & Obfuscate mail and host names\\
		File paths & Hide file paths in Linux and Windows format\\
		Executable files & Encode executable file names, e.g., py, sh, exe, jar\\
		All files & Hide all file names, ending with 2, 3, or 4 letters\\
		Keywords & Anonymize a list of custom keywords\\
		All & Apply all obfuscation methods listed above\\
		\bottomrule
	\end{tabular}
	}
	\label{tbl:anonym-tool-methods}
	\vcutM
\end{table}

We implement the unified trace format using the Parquet file format and the Snappy compression algorithm.
Parquet is a binary file format that is supported by many big data tools such as Apache Spark, Flink, Druid, and Hadoop~\cite{DBLP:conf/wosp/TalluriLAI19}. Many programming languages also have libraries to parse this format, such as PyArrow for Python and parquet-avro for Java.
Snappy\footnote{https://github.com/google/snappy} compression reduces the size of the dataset significantly and has low CPU usage during extraction.

Beside trace selection support and to address \wtareq{4}, the \gls{wta} offers several tools to facilitate and incentivize the continuous growth of the archive.
Most of these tools required significant engineering effort to develop, due to the typical challenges of big data processing (high volume, noisy data, diverse input-formats, etc.).
The \gls{wta} simplifies the upload of new traces by providing a set of parsing scripts for different trace sources, such as Google, Pegasus, and Alibaba.
Parsing traces can become non-trivial, once they grow both in complexity and size.
Such traces require big data tools, such as Apache Spark, and enough resources, a cluster, to compute.
Noisy data raise another non-trivial issue: both Google's and Alibaba's cluster data contained either anomalous fields, undocumented attributes, and non-DAG workflows. Some of these issues were never discovered by their respective communities and were corrected in our parsing tools.
Debugging, filtering, and correcting noisy big data requires significant compute power and detailed engineering.

Because traces may contain sensitive information, the \gls{wta} offers a \emph{trace anonymization tool}, which supports users to automatically replace privacy and security-related information, to avoid an accidental reveal of proprietary information.
Specifically, to remove sensitive information from trace files, we use two common techniques~\cite{clusterdata:Reiss2012}, \emph{culling} and \emph{transforming}.
Culling is done during trace conversion, by omitting parts of the raw trace data which do not match our workflow trace format.
For the transformation, as presented in Table~\ref{tbl:anonym-tool-methods}, our anonymization tool automatically scans the workflow trace file for sensitive data, such as IP addresses, file paths, names, etc., by string pattern matching. 
Beside these standard sensitive-data checks, the \gls{wta} offers the option to search for custom privacy-critical strings.

Finally, all matched strings are replaced by a salted SHA-256 hash key. 
This approach using cryptographic hash functions offers protection of sensitive data, while preserving the relationships between the matched values in the same trace file~\cite{clusterdata:Reiss2012}. 
Additionally, our tool hides potential relations to other trace files by adding a salt of length $16$ to the hash key generation, which is randomly generated on each tool run.

To validate traces, the \gls{wta} provides a \textit{validation} script that checks the integrity and summarizes important characteristics of a trace.
During trace conversion, using the validation script, we successfully identified several parse bugs and inconsistencies in the data that we subsequently corrected.

Specifically, because tasks build the base of each trace, our tool checks if all contained tasks are well defined. 
This, for example, means that all parsed control dependencies, such as children and parents, link only to existing tasks with valid properties. 
A task property is valid, if the parsed property type matches the property type definition, and the property value is allowed
e.g. task runtime $> 0$. 
Based on and similar to this fundamental validation, our tool provides options to check the workflow and data transfer properties to identify inconsistencies, as well.

These tools help combating perceived barriers to share data described by Sayogo et al.~\cite{sayogo2013exploring}.
Several technological barriers are addressed by using a unified format and validation (data architecture, quality, and standardization),
Legal and policy barriers are more difficult to address. 
Our anonymization tool aids in overcoming the data protection barrier, yet legal and other enforced policies may require tailored solutions.

Besides offering these tools, the \gls{wta} also hosts the trace data, addressing logistic and economical barriers.
The increasing focus on sharing data artifacts by the community, is lowering the barrier regarding competition for merit and reputation for quality and  bolsters the culture of open sharing.
Finally, each trace has its own DOI by also uploading it to Zenodo\footnote{https://zenodo.org/} which can be cited and thus provides authors with the appropriate credits (incentive barrier).
\vcutM

\begin{table*}[t]
	\centering
	\caption{Overview of the current \gls{wta} content, grouped by source. Legend: D=Domain, DS=Datasets, PA=parameters, PL=Platform, S=Setup, A=Applications, WL=workload, WF=workflow, T=task, U=user, G=group, *=minimum, Eng = Engineering, Sci = Scientific, Ind = Industry, TCH = Total Core Hours. Items in bold are workloads introduced by this work. Items where workflows are for the first time analyzed in this work are in {\it italics}. The symbol $\ddagger$ next to S7 indicates data with promise to release, but for which the legal forms have not been completed yet; \gls{wta} can already release all other workloads.} \label{tbl:current-content-summary}
	\vspace*{-0.35cm}
	\resizebox{\textwidth}{!}{
		\begin{tabular}{@{}lrlrrrrrrrrrrlr@{}@{}}
			\toprule
			Source ID. Name  &   \#WL    &   D  &   DS &   \#PA  &   \#PL    &   \#S   &  \#A   &   \#WF    &   \#T &   \#U    &   \#G &  Year(s) & Timespan & TCH\\ 
			\midrule
			\textbf{S1. Askalon Old} & 2  &  Eng   &   -  &   -  &   1    &    -  &   mixed  &  4,583 &   167,677  &   *7 &   *6 & 2007 &  19 months & 4,685,300\\
			\textbf{S2. Askalon New} & 67 &  Sci   &   -  &   *2  &   2    &    67  &   *3   &  1,835 &   91,599  &   *67 &   *67 & 2016 &  47 days & 193\\
			S3. LANL                 & 2  &  Sci   &   -  &   -  &   1    &    -   &   mixed  &  1,988,397 & 475,555,927  &   - &   - & 2011-2016 &  63 months & *9,625,431\\
			S4. \textit{Pegasus}              & 8  &  Sci   &   -  &   -  &   *6    &    -   &   8  &  56 &   10,573  &   9 &   - &  2011 &  4 days & 1,477\\
			\textbf{S5. Shell}       &	1 &	 Ind    &	-    &	-    &	1    &	-	&	mixed    &	3,403    &	10,208    &	-    &	-  & 2016  &	10 minutes & 25\\
			\textbf{S6. SPEC}	     &	2 &	 Sci	&	-    &	-    &	1    &	-	&	mixed    &	400    &	28,506    &	-    &	- & 2017 & - & 1,231\\
			S7. Two Sigma$^\ddagger$         &	2 &	 Ind	&	-    &	-    &	1    &	-	&	mixed    &	41,607,237    &	50,518,481    &	610    &	1  & 2016 &	16 months & 69,992,196\\
			S8. \textit{WorkflowHub}	         &	10	&	Sci	&	*5	&	*4	&	5	&	-	&	3	&	10	&	14,275	&	10	&	-	& 2017 & - & 52\\
			S9. \textit{Alibaba}	         &	1	&	Ind	&	-	&	-	&	1	&	-	&	mixed	&	4,210,365	& 1,356,691,136 &	1	&	1 & 2018	&	8 days & 1,526,925,484\\
			S10. Google	         &	1	&	Ind	&	-	&	1	&	1	&	-	&	mixed	&	494,179	& 17,810,002 &	430	&	1	& 2011 &	29 days & 434,821,345\\ \midrule
			Total & 96 & - & *5 & *7 & *20 & 67 & - & 48,310,465 & 1,900,898,384 & *1,134 & *76 & - & - & 2,046,052,734 \\
			\bottomrule
		\end{tabular}
	}
\vcutM
\end{table*}

\begin{table*}[t]
\caption{Overview of properties available per source. Legend: $\checkmark$ = available, $\sim$ = partially available, blank = not available, Task details = individual task information.}\label{tbl:trace-properties-per-source}
\resizebox{\textwidth}{!}{
\begin{tabular}{@{}lcccccccc@{}}
\toprule
Source ID & Task details & Task resource req. & Structural information & Disk & Memory & Network & Energy & \glspl{nfr} \\ \midrule
S1        & $\checkmark$      & $\checkmark$    & $\checkmark$      &   &   &   &  &   \\
S2        & $\checkmark$      & $\checkmark$    & $\checkmark$      &   &   &   &  &   \\
S3        & $\sim$ & $\checkmark$    & $\sim$ &   &   &   &  & $\checkmark$    \\
S4        & $\checkmark$      & $\checkmark$    & $\checkmark$      &   &   &   &  &   \\
S5        & $\checkmark$      & $\checkmark$    & $\checkmark$      &   &   &   &  &   \\ \midrule
S6        & $\checkmark$      & $\checkmark$    & $\checkmark$      &   &   & $\checkmark$    &  &   \\
S7        & $\checkmark$      & $\checkmark$    & $\sim$ &   & $\checkmark$    &   &  &   \\
S8        & $\checkmark$      & $\checkmark$    & $\checkmark$      &   & $\checkmark$    & $\checkmark$    &  &   \\
S9        & $\checkmark$      & $\checkmark$    & $\checkmark$      & $\checkmark$    & $\checkmark$    & $\checkmark$    &  &   \\
S10       &     & $\checkmark$    & $\sim$ & $\checkmark$    & $\checkmark$    &   &  & $\checkmark$    \\ \bottomrule
\end{tabular}
}
\end{table*}

\subsection{Current Content}
\label{ssct:current-content}

Having a diverse set of traces available is necessary to use in experimentation.
When using traces in experimentation, different traces should be used to prove generality of the proposed approach (see Section~\ref{sct:validation-and-threats}).
Gathering and parsing raw logs and other traces requires significant computing effort.
Using 16 nodes (32 eight-core Xeon E5-2630 v3 and 1TB RAM) from the Dutch DAS5 super computer~\cite{bal2016medium}, several traces require up to a day to compute using big data tools such as Apache Spark.
In total, the WTA team spent more than two person months on converting traces to the unified trace format.
By offering these parse scripts and the data, we contribute to \wtareq{4}.

The \gls{wta} features currently $96$ workloads from 10 different sources, with over $48$ million workflows and $2$ billion CPU core hours.
Each workload is uniquely identified by a combination of the following properties if available: source, runtime environment, application, and application parameters~\cite{DBLP:conf/europar/0002SDL16}.
Table~\ref{tbl:current-content-summary} and~\ref{tbl:trace-properties-per-source} summarize these traces. %
From these tables we observe that \gls{wta} contains a vast amount of different traces, from different sources and domains, with various number of workflows, properties, number of tasks, timespans, and core hour counts.
Although supported by our format, no trace currently has information on energy consumption, highlighting the need of such traces~\cite{da2014community}.
These traces are collected by combining open-access data (logs, traces, etc.) and closed-access data throughout the years in collaboration with both industry and academia.
This contributes to \wtareq{1}. %

This diversity enables new workflow management techniques and systems to be thoroughly tested for their feasibility, strengths, and, equally important, weaknesses.

An example of the usability of different traces is in emerging fields.
Serverless is a new and emerging field, hence no traces are available yet.
One of our collaborators used the Shell (IoT) workload, which features small and short-running workflows, much alike serverless functions, to experiment with a serverless workflow engine.
Additionally, from these traces additional insights can be gathered regarding task and job runtimes, sizes, arrival patterns, resource consumption, etc.

We encourage the community to continuously contribute traces to the archive to improve coverage of domains, maintain its representativeness of the jobs being executed in real-world by both academia and industry, offer more artifacts to prove robust performance of e.g., policies, and offer more diversity in trace statistics.
By making available methods for contribution, we fulfill \wtareq{5}. 
\vcutS

\section{A Characterization of \\Workloads of Workflows}
\label{sct:workflow-characterization}\label{sec:characterization}

\begin{table*}[t]
	\centering
	\caption{The design and setup of our characterization.}
	\label{tbl:characterization-experiments-setup}
	\vspace*{-0.35cm}
	\adjustbox{max width=\linewidth}{
	\begin{tabular}{@{}llllll@{}}
		\toprule
		ID & \S & Description                                               & Traces                               & Metric                & Granularity    \\ \midrule
		E1 & \ref{ssct:exp_structual_patterns} & Analyze structural patterns in workflows per domain & All but S3, 7, 10 & Structural patterns   & Workflow level \\
		E2 & \ref{ssct:exp_longitudinal_trend} & Longitudinal analysis & S1, S3, S7, S9, S10 & Tasks per day & Workload level \\
		E3 & \ref{ssct:burstiness} &  Analysis of burstiness per trace & All but S4-8  & Hurst exponent & Workload level \\
		E4 & \ref{ssct:exp_level_of_parellelism} &  Measure the level of parallelism per workflow & All but S3, 7, 10 & Level of parallelism  & Workflow level \\
		E5 & \ref{ssct:exp_cp_length} &  Analysis of critical path length & All but S3, 7, 10 & Critical path length & Workflow level \\
		E6 & \ref{ssct:exp_cp_runtime} &  Analysis of critical path runtime & All but S3, 7, 10 & Critical path runtime & Workflow level \\
		E7 & \ref{ssct:exp_task_interarrival_times} &  Task interarrival time analysis & All & Interarrival time & Workload level \\
		 \bottomrule
	\end{tabular}
}
\vcutL
\end{table*}

To answer \resquestion{3}, we perform in this section a characterization of the workloads in the \gls{wta}.
We characterize workloads using a variety of metrics and properties, including workflow size, resource usage, and structural patterns.
Our characterization reveals significant differences between workloads from different domains and sources. Such differences further support our claim that the community needs to look beyond just scientific workloads, and consider a wider range of domains and sources for experimental studies.

\vcutS

\subsection{Structural Patterns}
\label{ssct:exp_structual_patterns}

\begin{figure}[t]
	\centering
	\includegraphics[width=0.9\columnwidth]{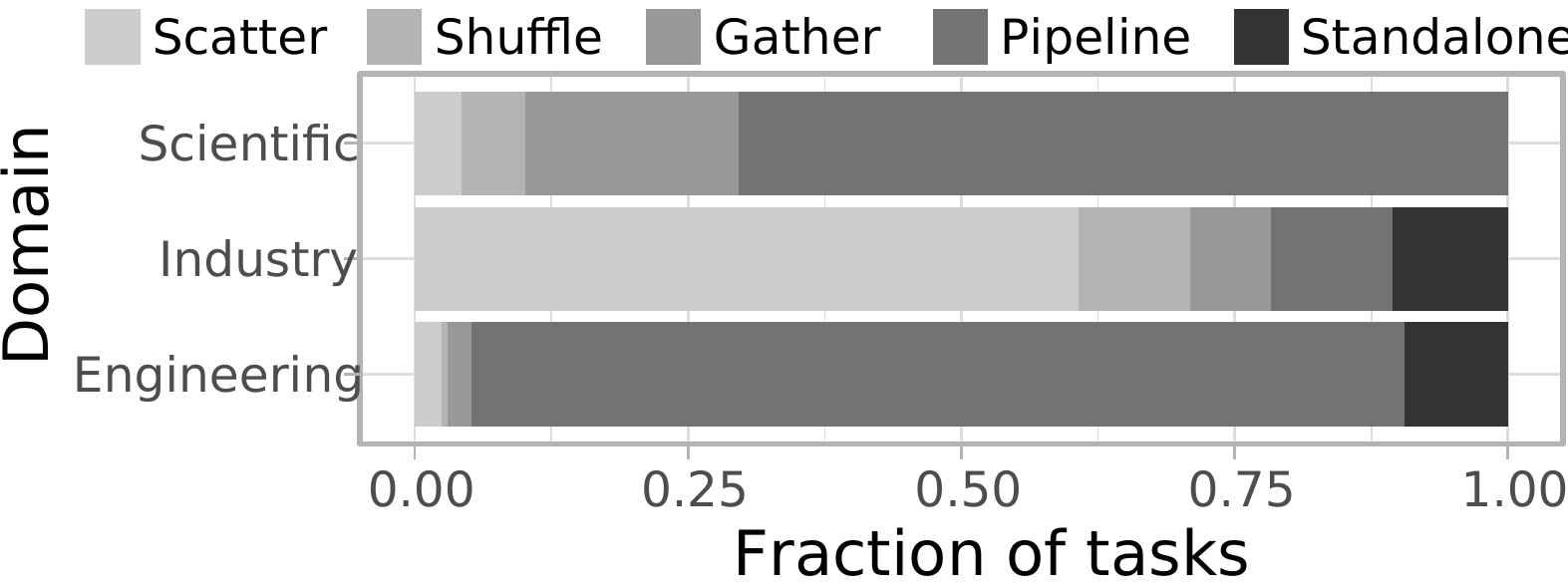}
	\vcutM
	\caption{Structural workflow patterns, per domain.} \label{fig:exp_structual_patterns}
    \vcutL
\end{figure}

\begin{description}
\maindfinding{mf:strucural-variety}{Scientific, industrial, and engineering workflows exhibit various structural patterns, but at least 60\% of tasks in a domain match the dominant pattern of that domain.}
\maindfinding{mf:ind-scatter}{Industry workflows stand out by exhibiting primarily scatter patterns, as opposed to pipeline operations.}
\end{description}

This characterization quantifies five structural patterns in workflows often used by researchers~\cite{bharathi2008characterization}: scatter (data distribution), shuffle (data redistribution), gather (data aggregation), pipeline, and standalone (process).
Investigating these structural patterns is important to understand the types of applications being executed and tune a system's performance.
We exclude from this analysis the LANL, Two Sigma, and Google traces, which lack structural information, that is, task parent-child relationship information.

Figure~\ref{fig:exp_structual_patterns} depicts the structural patterns found per domain.
From this figure, we observe that in each domain a dominant pattern emerges that accounts for 61--85\% of tasks.
In the scientific and engineering domains, the majority of tasks are simple pipelines. 
Interestingly, the industrial workflows include primarily scatter operations.
This observation matches known properties of the Alibaba trace, which accounts for over 99\% of tasks with structural information we analyzed in this domain.
In particular, the Alibaba trace includes MapReduce jobs, each consisting of many ``map'' tasks (scatter operations) and a smaller number of ``reduce'' tasks (gather operations).
\vcutS

\subsection{Arrival Patterns}
\label{ssct:exp_longitudinal_trend}

\begin{figure}[t]
	\centering
	\includegraphics[width=0.9\columnwidth]{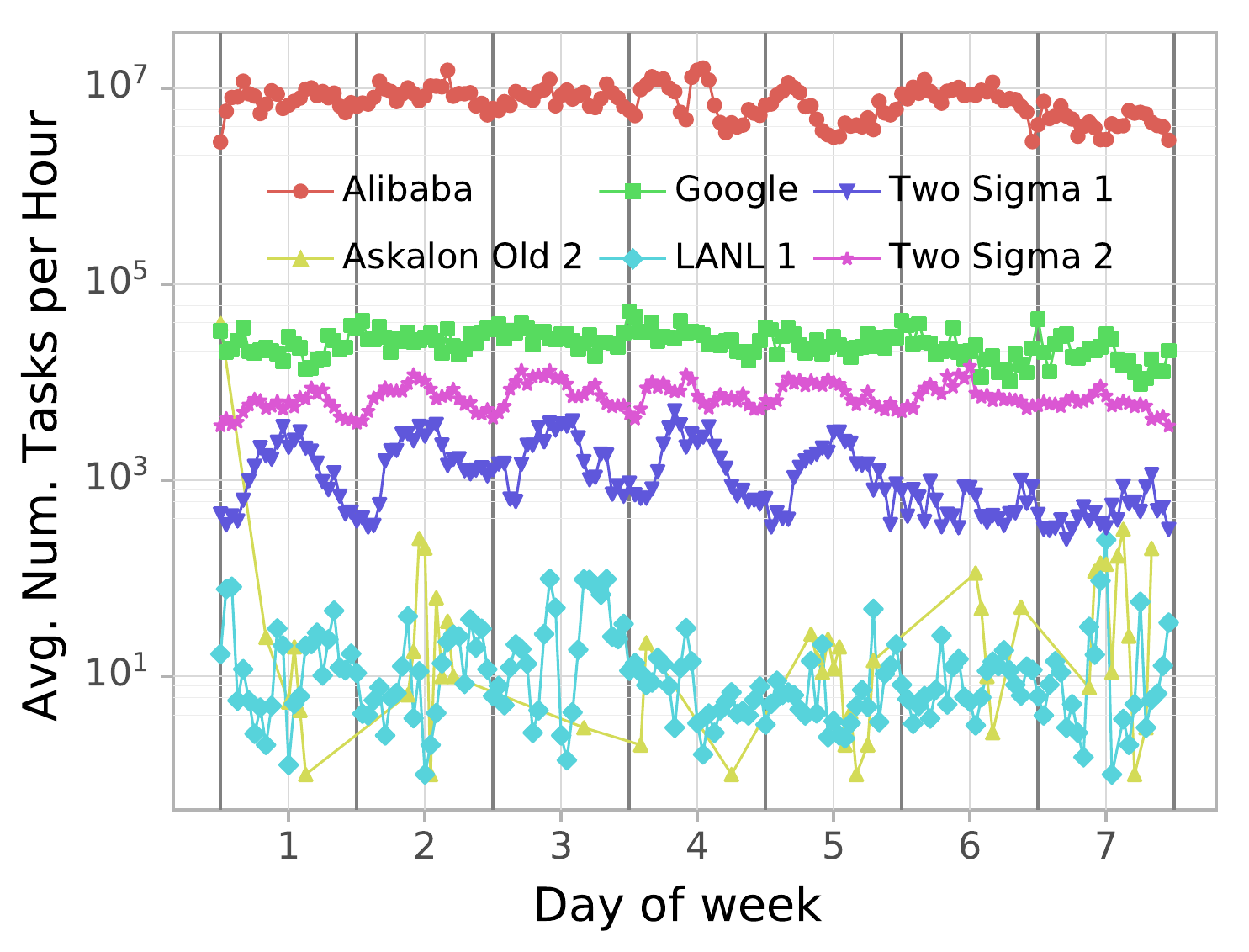}
	\vcutL
	\caption{Daily task-arrival trend, per source.} %
	\label{fig:exp_longitudinal_trend}
    \vcutL
\end{figure}

\begin{description}
	\maindfinding{mf:industrial-arrival-rate}{From all domains, industrial traces show on average orders of magnitude higher rates of task arrival.}
	\maindfinding{mf:scientific-high-variability}{Scientific traces can show high variability in task arrival rates, unlike industrial and engineering traces.}%
	\maindfinding{mf:two-sigma-workday-pattern}{Two Sigma shows a typical workday diurnal pattern.}
\end{description} 

To investigate the weekly trends that may appear in workload traces, we depicts in Figure~\ref{fig:exp_longitudinal_trend} for several traces the average number of tasks that arrive per day of the week.
We omit the Askalon new source from the hourly task-arrival plot as they contain 4 or 5 data points, which is too few to plot a trend.
We observe that traces have significantly different arrival rates and patterns.
The Alibaba trace features the highest task arrival rates, peaking at over 10,000,000 tasks per hour. Google and the Two Sigma workloads follow with 10-10,000 tasks per hour.
This shows that industrial workloads included in this work have significantly more tasks per hour than the other compute environments, which agrees with companies such as Alibaba and Google operating at a global scale.
Also interesting to notice is that besides Askalon Old 1, the non-industrial traces show significant fluctuations throughout the week, whereas both Alibaba and Google do not.
This might be due to the global, around-the-clock operation of Alibaba's and Google's services, which can lead to a more stable task arrival rate.

\begin{figure}[t]
	\centering
	\includegraphics[width=0.9\columnwidth]{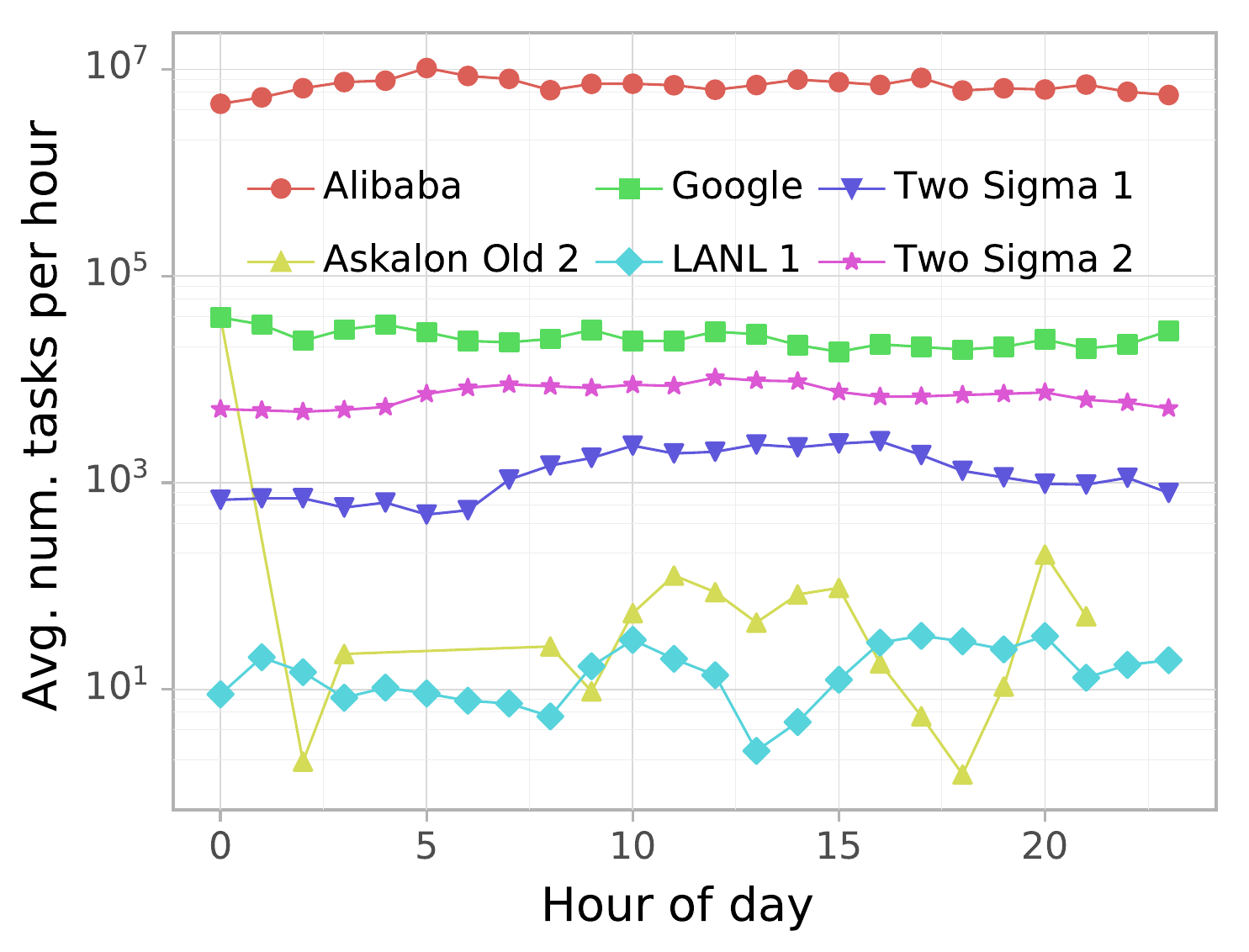}
	\vcutL
	\caption{Hourly task-arrival trend, per source.} %
	\label{fig:exp_longitudinal_trend_daily}
\vcutL
\end{figure}

To observe differences in daily trends, we depict the average task rate per hour of day in Figure~\ref{fig:exp_longitudinal_trend_daily}.
This figure reaffirms our observation that the two largest traces--Alibaba and Google--and the Askalon Old 1 trace have a relatively stable arrival pattern throughout the day.
In contrast, the Two Sigma 1 trace exhibits a typical office hours pattern; task arrival rates increase around hour 7 and start dropping around 17. The same pattern occurs to a lesser extent in the Two Sigma 2 trace.
The highly variable arrival rates of tasks in the LANL traces, as observed in Figure~\ref{fig:exp_longitudinal_trend}, are also evident in our analysis of daily trends. We study this in more depth in Section~\ref{ssct:burstiness}.

\subsection{Burstiness}
\label{ssct:burstiness}

\begin{figure}[t]
	\centering
	\includegraphics[width=0.9\columnwidth]{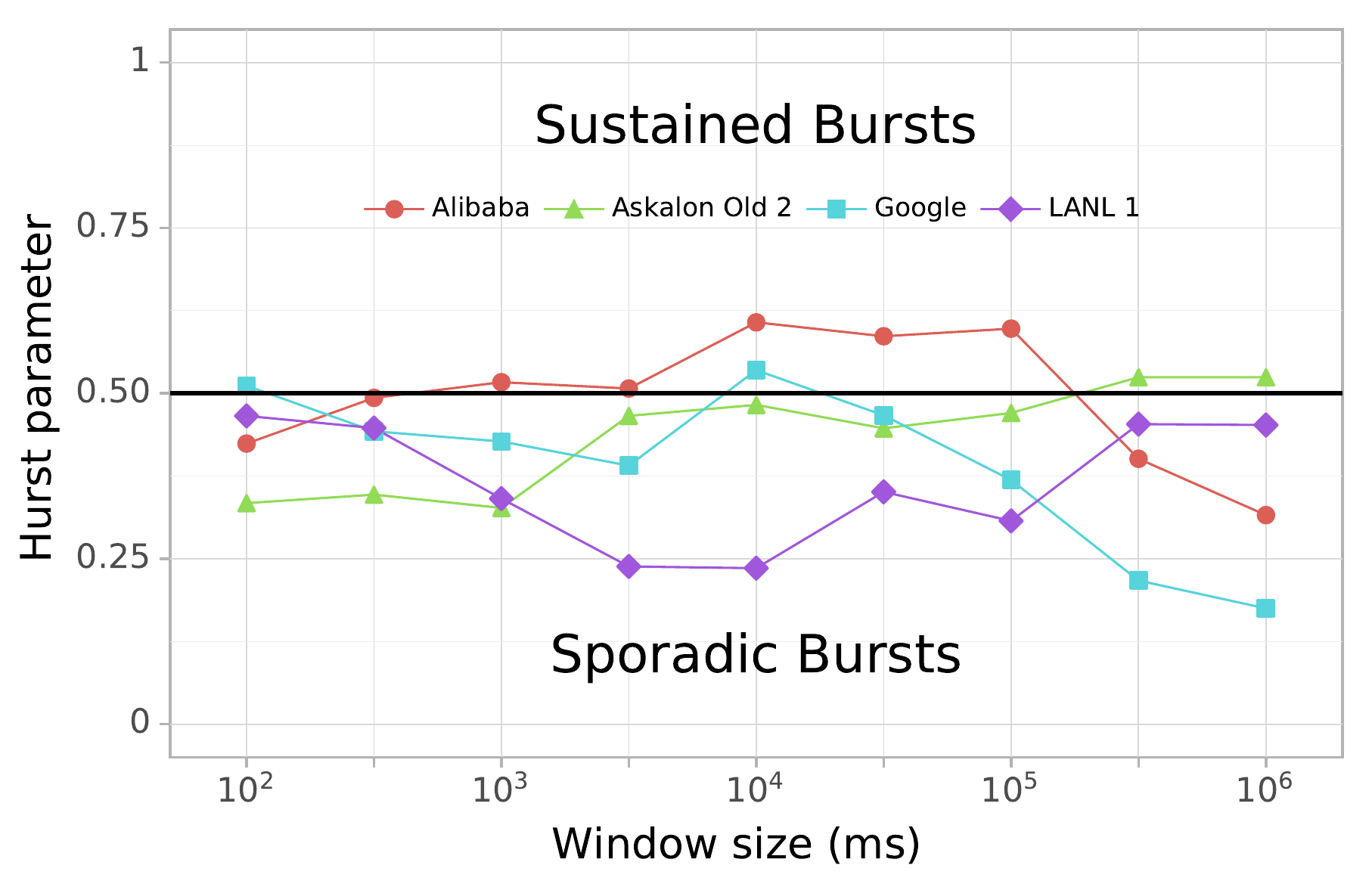}
	\vcutL
	\caption{Hurst exponent estimations for different time windows per trace. Horizontal axis does not start at zero.} 
	\label{fig:exp_bustiness}
	\vcutL
\end{figure}

\begin{description}
	\maindfinding{mf:bursty-small-windows}{Most traces investigated exhibit bursty behavior within small window sizes.}
	\maindfinding{mf:lanl-bursty}{The LANL trace exhibits maximum burstiness at medium window sizes.}
	\maindfinding{mf:alibaba-google-bursty}{The largest traces (Alibaba and Google) exhibit uniquely bursty behavior: low burstiness at small and high burstiness at large, window sizes.}
\end{description}

To investigate if workloads expose bursty behavior, a special kind of arrival pattern, the Hurst exponent $H$ is used. 
$H$ quantifies the effect previous values have on the present value of the time series.
A value of $H < 0.5$ indicates a tendency of a series moving in the opposite direction based on the previous values, and thus exhibit jittery behavior (sporadic burst).
A value of $H > 0.5$ indicates a tendency to move in the same direction, and thus towards well defined peaks (sustained burst).
When $H = 0.5$, the series behaves like a random Brownian motion.

In this experiment, we inspect busty behavior by computing the Hurst exponent for task arrivals.
The results of this experiment are visible in Figure~\ref{fig:exp_bustiness}.
From this figure, we observe most traces depict bursty behavior at least for one of small, medium, and large window size. They are also not bursty for at least one window size.
This is expected, as in most systems task arrivals vary at (sub-)second interval.
Interestingly, LANL traces exhibit most bursty behavior at medium window sizes.
This might be due to national laboratories workflows being submitted in waves. A wave is submitted all at once leading to a burst. But, a wave itself is processed smoothly. The workload is also stable over longer time periods as evidenced by $H \approx 0.5$ for larger windows.
Finally, the two largest traces in this work, Alibaba and Google, exhibit increasingly burst behavior for larger windows.
This indicates that for larger arrival times, the workloads (in absolute numbers) vary more than for the other sources. This matches the observations in Section~\ref{ssct:exp_longitudinal_trend}.

\subsection{Parallelism in Workflows}
\label{ssct:exp_level_of_parellelism}

\begin{figure}[t]
	\centering
	\includegraphics[width=\columnwidth]{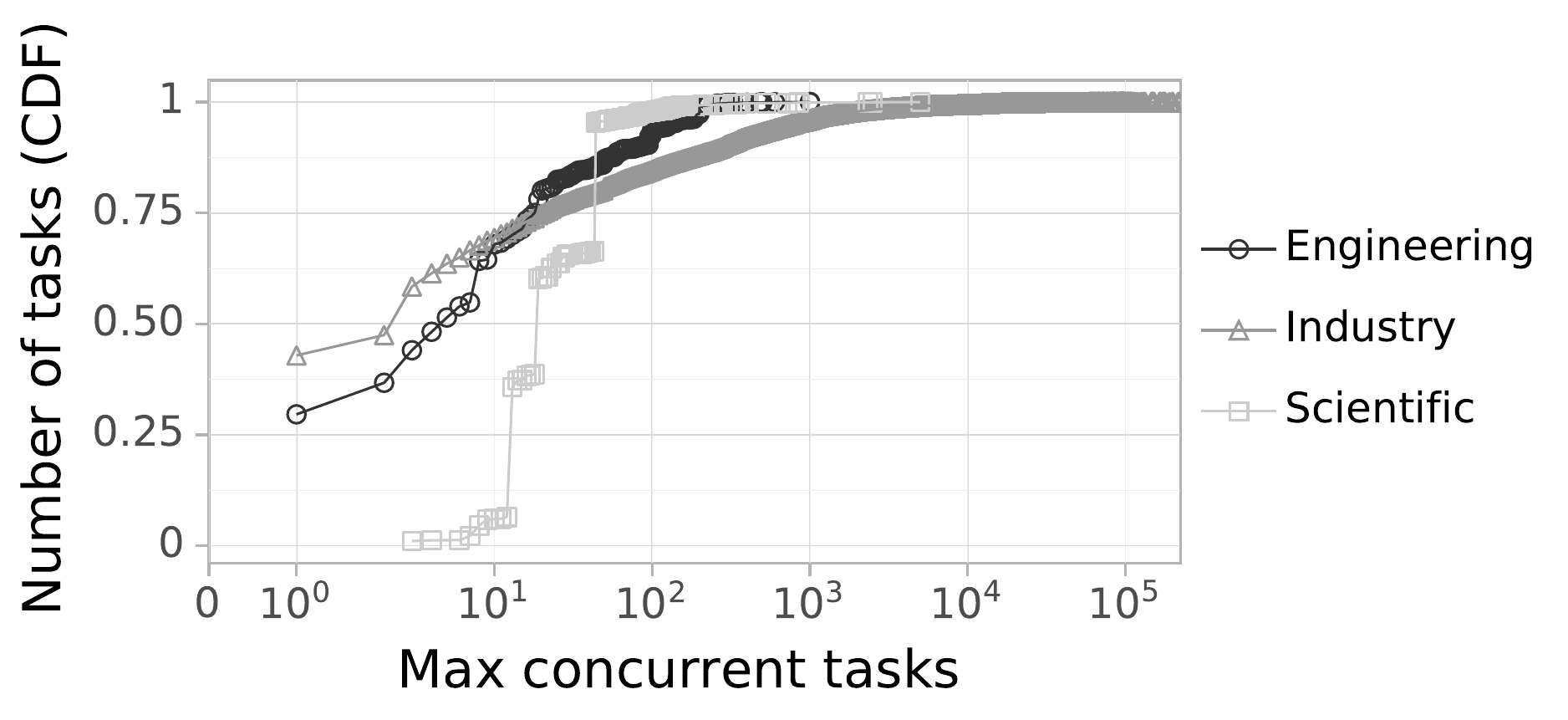}
	\vcutXL
	\caption{Workflow level-of-parallelism, per domain.}
	\label{fig:exp_level_of_parellelism}
    \vcutL
\end{figure}

\begin{description}
	\maindfinding{mf:parallelism-per-domain-differs}{Task parallelism per workflow can differ significantly between workload domains and sources.}
	\maindfinding{mf:ind-highest-parallelism}{Industrial workflows exhibit the highest level of parallelism.}
	\maindfinding{mf:alibaba-highest-parallelism}{Out of all sources, Alibaba workflows have the highest level of parallelism, followed by Pegasus and WorkflowHub.} %
\end{description} 

With the structural patterns observed, we investigate if the large occurrence of the pass-through patterns expresses in a high level of parallelism.
The level of parallelism indicates how many tasks can maximally run in parallel for a given workflow, provided sufficient resources.
Figure~\ref{fig:exp_level_of_parellelism} depicts the approximated level of parallelism per domain.
The approximation algorithm used produces results very close to the true level of parallelism as demonstrated by Ilyushkin et al.~\cite{ilyushkin2015scheduling}.
From this figure, we observe the industrial domain exhibits the highest level of 99th percentile parallelism, up to hundreds of thousands of tasks.
This is likely a consequence of the many MapReduce workflows, which are highly-parallel by nature, that are present in the Alibaba trace.
Alibaba also contains bag of tasks workflows, which by nature have a high parallelism.%
Scientific workflows exhibit low median parallelism but high 99th percentile parallelism, featuring levels of parallelism up to thousands of tasks.
Engineering traces exhibit a moderate amount of median parallelism, between industry and scientific, with at most 1000 concurrent running tasks.

\begin{figure}[t]
{
	\centering
	\includegraphics[width=\columnwidth]{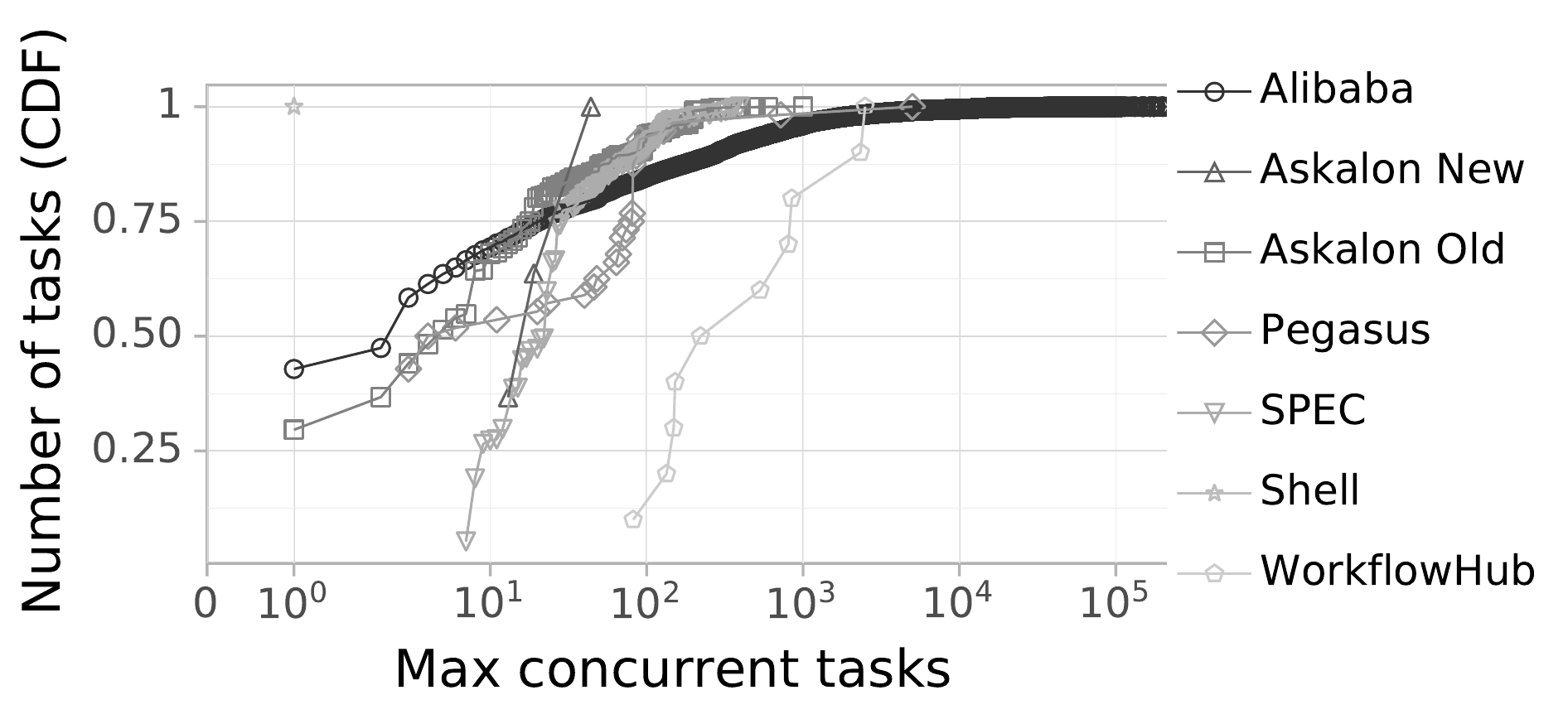}
}
	\vcutXL
	\caption{Workflow level-of-parallelism, per source. %
	Curves are %
	shaded by domain, to further reveal patterns.}
	\label{fig:exp_level_of_parellelism_source}
	\vcutL
\end{figure}

Figure~\ref{fig:exp_level_of_parellelism_source} shows the level-of-parallelism per source.
From this figure, we observe that Alibaba exhibits the highest levels of parallelism, as discussed previously.
Second are the Pegasus and WorkflowHub workflows.
These sources contain a variety of scientific applications, commonly known for their parallel structures, as observed in Section~\ref{ssct:exp_structual_patterns}.
Other traces demonstrate less parallelism, with up to 100 concurrent running tasks.
As Shell exist entirely of sequential pipelines, the source does not exhibit any variation.

\vcutS

\subsection{Limits to Parallelism in Workflows}
\label{ssct:exp_cp_length}

\begin{description}
	\maindfinding{mf:cp-differs-per-domain}{Workflows from the scientific domain have significantly different critical-path lengths.}
	\maindfinding{mf:eng-highest-cp}{The amount of tasks on the critical path is the highest for engineering workflows.}
    \maindfinding{mf:ind-cp-higher-sci}{Although highly parallel, industrial workflows exhibit longer critical paths than scientific workflows.}
\end{description}

\begin{figure}[t]
	\centering
	\includegraphics[width=\columnwidth]{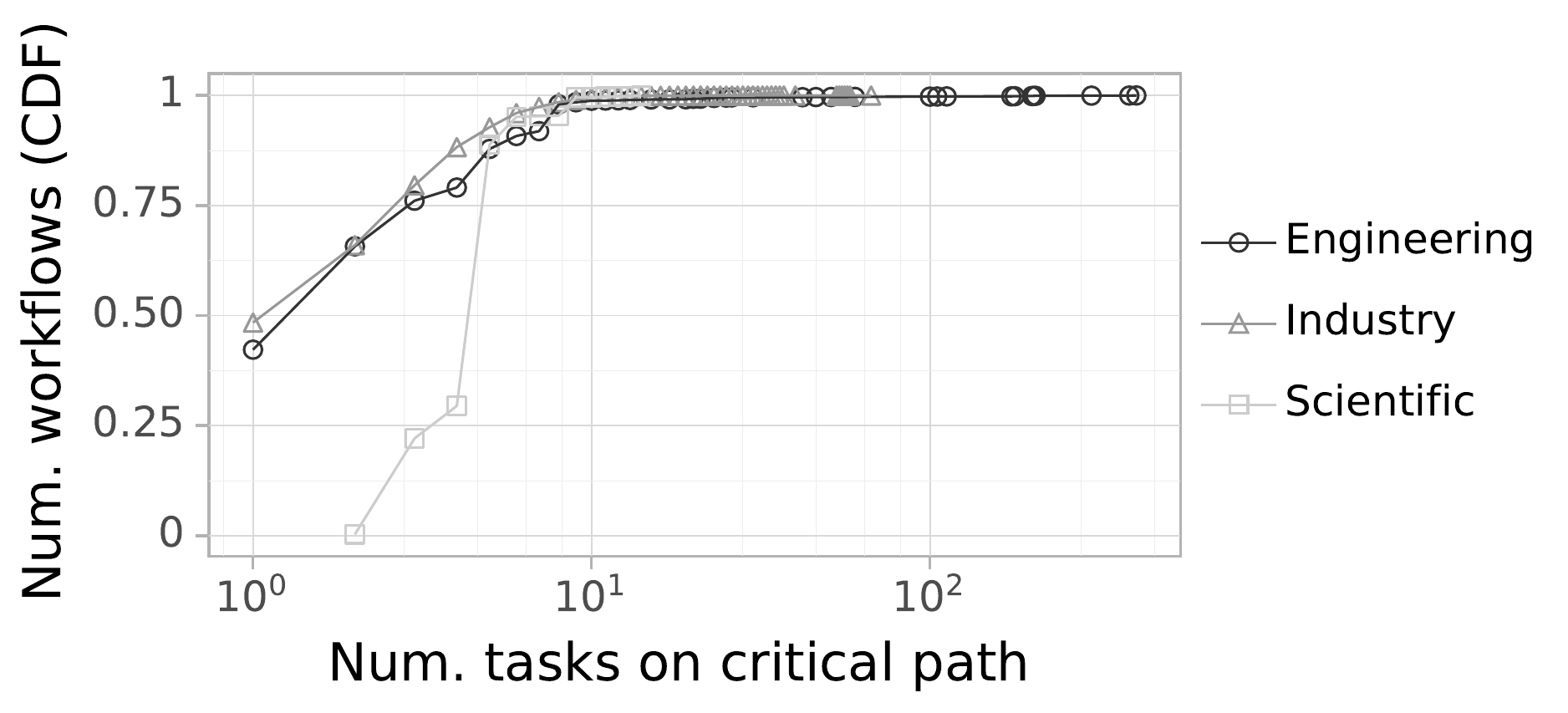}
	\vcutXL
	\caption{Workflow \acfp{cp}, per domain.}
	\label{fig:exp_cp_length}
	\vcutL
\end{figure}

The \gls{cp} refers to the longest sequence of dependent tasks in a workflow, from any entry task to any exit task. 
By quantifying the \gls{cp} length, we investigate if workflow runtimes are primarily dominated by a few heavy tasks, or by many small tasks.
Figure~\ref{fig:exp_cp_length} presents the results of this characterization per workload domain.
From this figure we observe the \gls{cp} length for engineering workflows is the highest. 
This matches with the parallelism observations in Sections~\ref{ssct:exp_structual_patterns} and~\ref{ssct:exp_level_of_parellelism}.
Interestingly, even though industrial workflows are often highly parallel, their critical paths are often longer than those of scientific workflows.
This indicates that industrial workflows are bigger in size than scientific workflows, which our data supports.

\begin{figure}[t]
	\centering
	\includegraphics[width=\columnwidth]{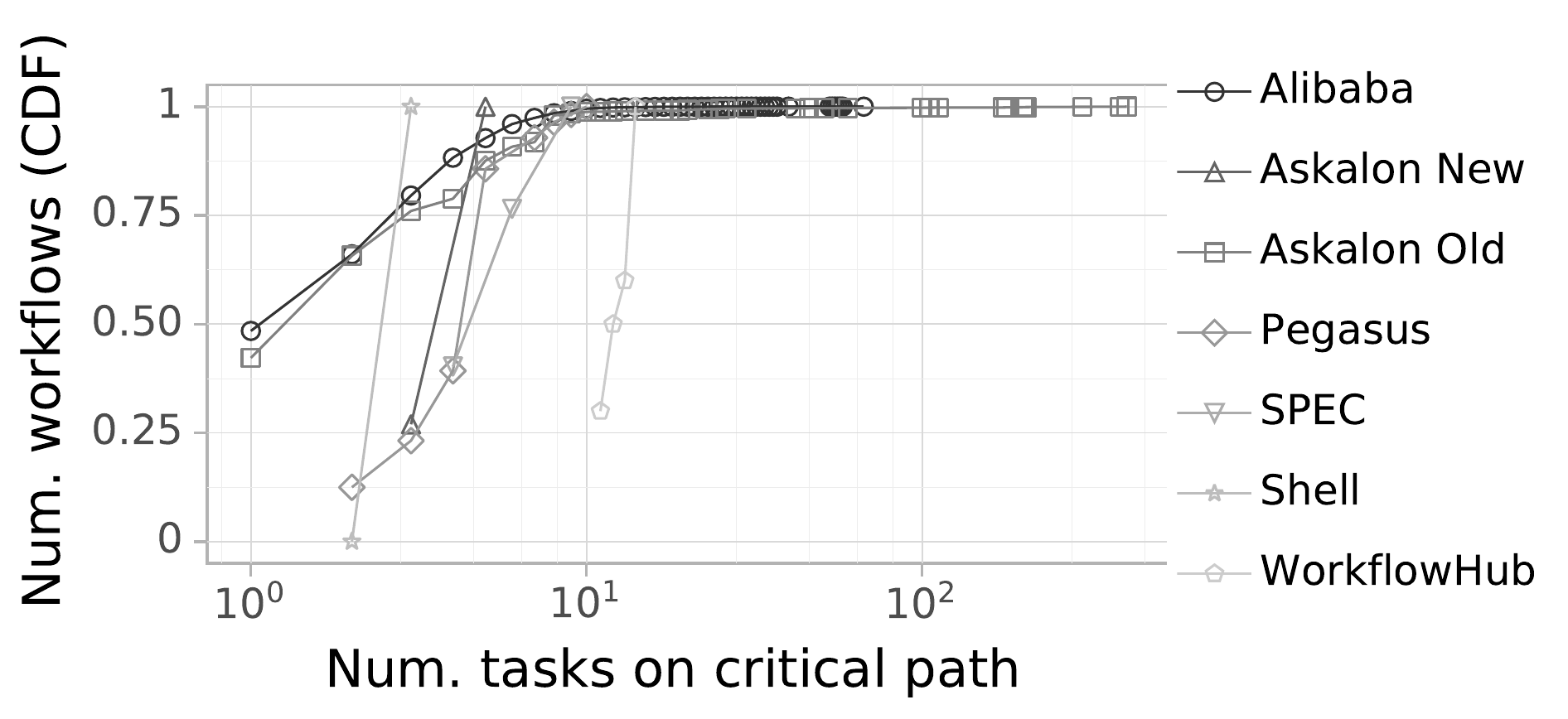}
	\vcutXL
	\caption{Workflow \acfp{cp}, per source.
	Curves are %
	shaded by domain, to further reveal patterns.}
	\label{fig:exp_cp_length_source}
	\vcutL
\end{figure}

Figure~\ref{fig:exp_cp_length_source} presents the results of \gls{cp} characterization per workload source.
From this figure we observe the \gls{cp} length differs significantly per trace.
Based on the prior findings, the engineering traces are expected to show longer critical paths.
As we can observe, the Askalon old traces contains workflows with the longest critical path.
Alibaba workflows also exhibit long critical paths, indicating their workflows next to being highly parallel, also contain a lot of tasks with stages.
More concentrated, the other traces exhibit lower critical path lengths, yet the traces are still clearly distinct.
As the Shell trace contains solely sequential workflows, the critical path length is one.

\subsection{Critical Path Runtime}
\label{ssct:exp_cp_runtime}

\begin{description}
	\maindfinding{mf:eng-highest-runtime}{Engineering workflows have the highest critical path runtime.}
    \maindfinding{mf:ind-runtime-higher-sci}{Although highly parallel, industrial workflows exhibit longer runtimes than scientific workflows.}
\end{description}

\begin{figure}[t]
	\centering
	\includegraphics[width=\columnwidth]{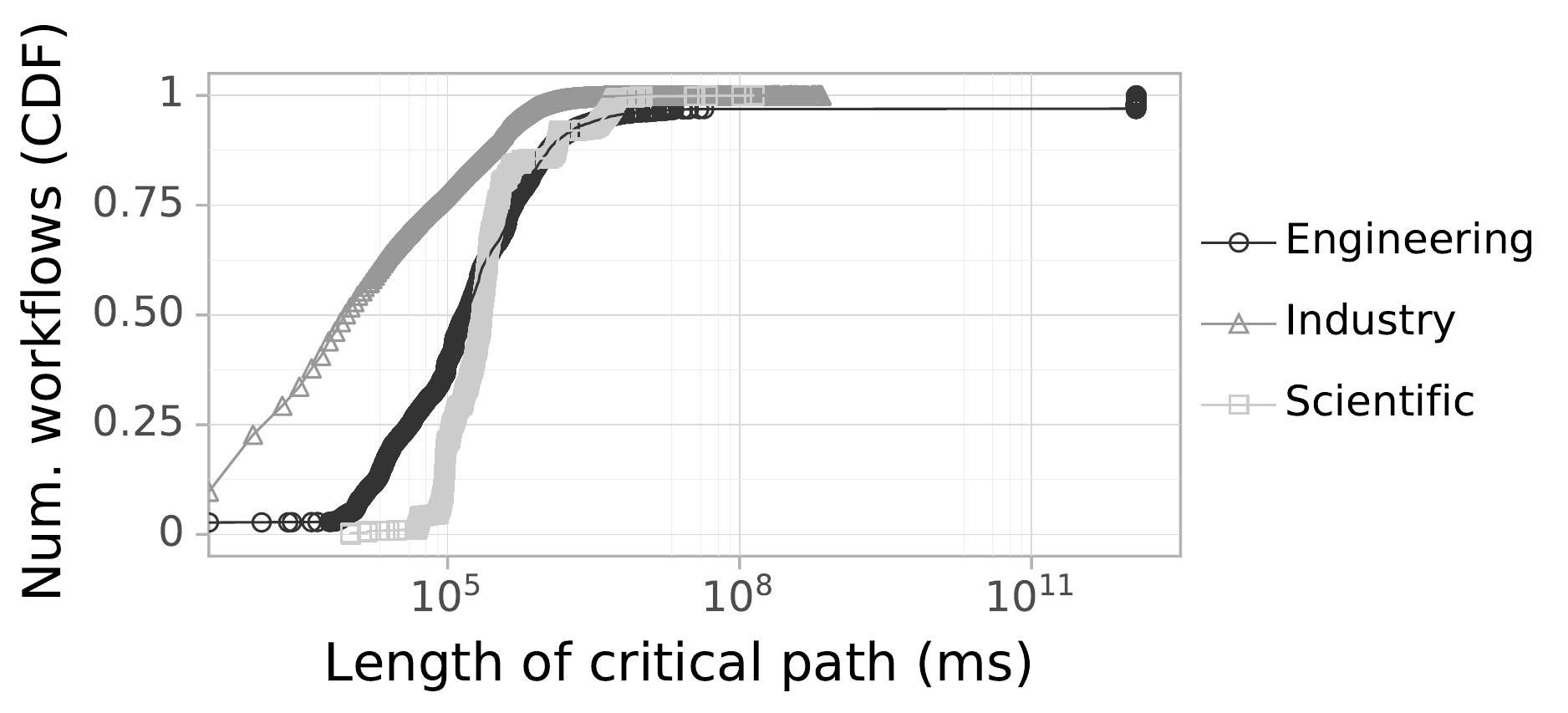}
	\vcutXL
	\caption{Workflow critical path runtime, per domain.}
	\label{fig:exp_cp_runtime}
	\vcutL
\end{figure}

The critical path runtime is the sum of runtimes of all tasks on the critical path. It is the minimum amount of time required to run a workflow on a given infrastructure.
Figure~\ref{fig:exp_cp_runtime} represents presents the results of this characterization per workload domain. We observe that the results are very similar to those obtained by characterizing the critical path length, per domain, in Section~\ref{ssct:exp_cp_length}. Engineering workflows have the highest critical path runtime, followed by industry and scientific workflows.

\begin{figure}[t]
	\centering
	\includegraphics[width=\columnwidth]{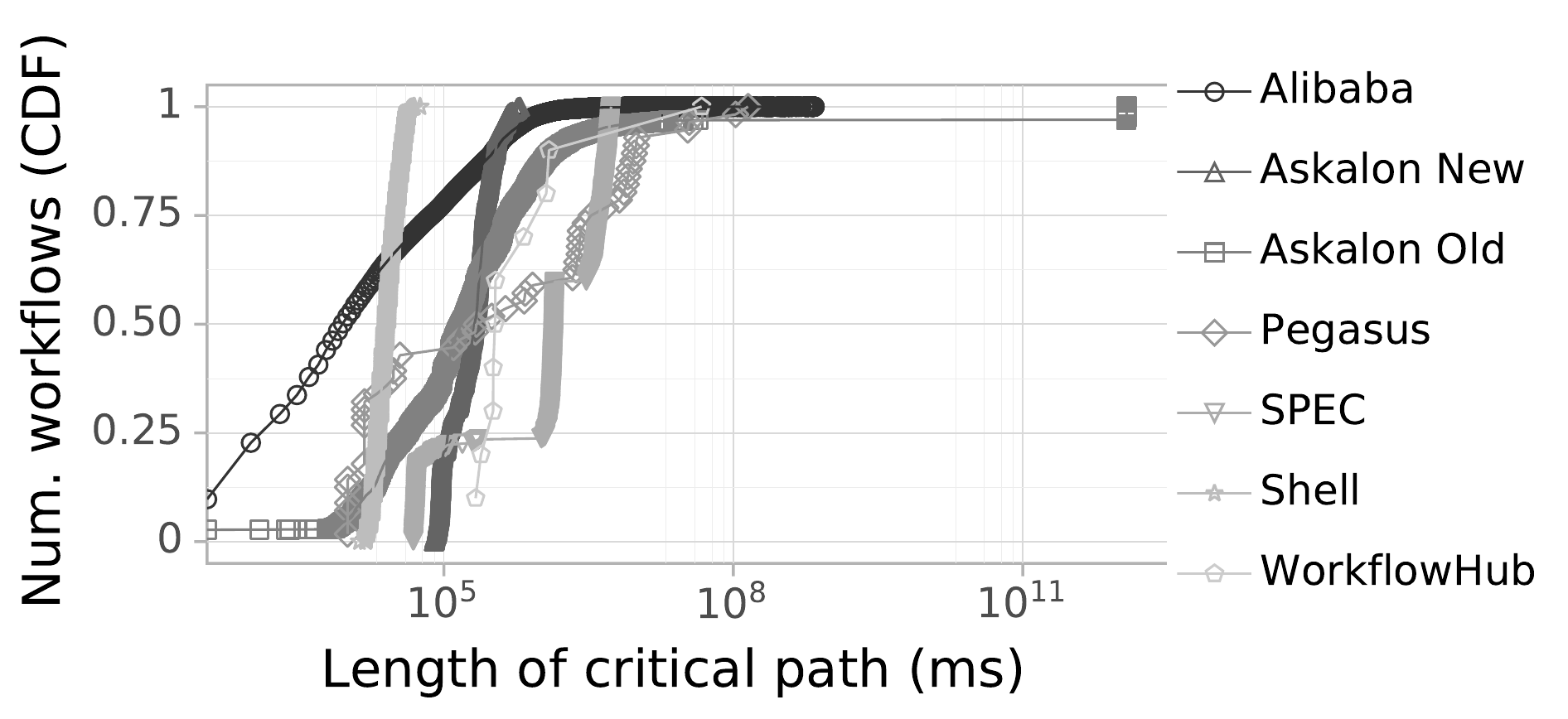}
	\vcutXL
	\caption{Workflow critical path runtime, per source.}
	\label{fig:exp_cp_runtime_source}
	\vcutL
\end{figure}

Figure~\ref{fig:exp_cp_runtime_source} presents the results of critical path runtime characterization per workload source. We observe that the results are very similar to those obtained by characterizing the critical path length, per source, in Subsection~\ref{ssct:exp_cp_length}. Askalon Old has the highest runtime followed by Alibaba. The runtimes distributions of individual sources are clearly distinguishable.

\subsection{Task Interarrival times}
\label{ssct:exp_task_interarrival_times}

\begin{description}
	\maindfinding{mf:industry-lowest-task-intararrival}{Almost all tasks in the industrial domain arrive within milliseconds after one-another.}
	\maindfinding{mf:industry-highest-task-intararrival}{The industrial domain also exhibits the longest task interarrival times.}
	\maindfinding{mf:engineering-lowest-interarrival-times}{The engineering domain exhibits the highest spread of task interarrival times.}
	\maindfinding{mf:askalon-and-lanl-lowest}{Per source, the Askalon (old and new) and LANL exhibit the lowest task interarrival times.}
	\maindfinding{mf:two-sigma-longest-interarrival-times}{Two Sigma features the longest task interarrival times, possibly indicating downtime.}
\end{description}

In this experiment, we inspect the task interarrival times.
Traces not containing task-level information, such as one of the LANL traces, are omitted.
By quantifying the task arrival times, we gain insight into the speed scheduler in different environments must make decisions.
Low task interarrival times translate to systems with a continuous stream of incoming tasks, possibly at a high rate.
In such systems, schedulers must make often sub-second decisions to avoid tasks delaying, which could have important implications on \gls{qos} metrics such as job makespan and task turnaround time.
High task interrival times translate to systems with longer periods between the arrival of two consecutive tasks.
However, this does not mean the amount of tasks arriving might be low.
In situations where bags-of-tasks are submitted simultaneously, the task-interarrival times between these tasks will be zero, yet the time between the arrival of two bags-of-tasks may be long.

\begin{figure}[t]
	\centering
	\includegraphics[width=0.9\columnwidth]{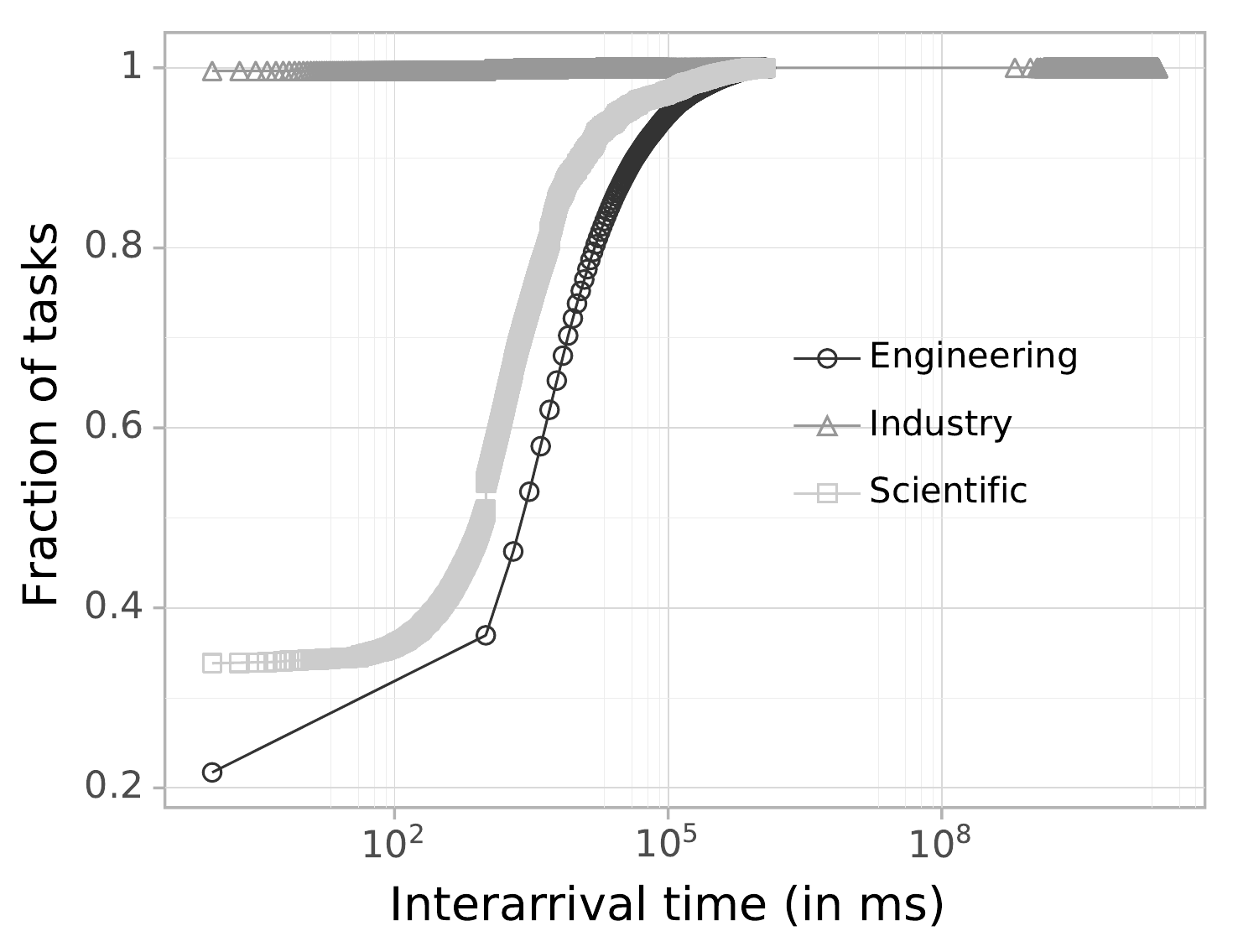}
	\vcutL
	\caption{CDF of task interarrival times, per domain.} %
	\label{fig:exp_task-interarrival-time_domain_trend}
	\vcutL
\end{figure}

Figure~\ref{fig:exp_task-interarrival-time_domain_trend} depicts the CDF of task interarrival times per domain.
From this figure we observe that almost all tasks in the industrial domain have a task interarrival time of less than 10 milliseconds.
This means industrial task schedulers must make decisions at the millisecond level.
Interestingly, the industrial domain also exhibits the highest task interarrival times.

The scientific domain shows roughly 30\% of all tasks arrive within 10 milliseconds of one another.
Roughly 93\% of all tasks have an interarrival time of less than 100 milliseconds.

For the engineering domain, roughly 21\% of all tasks have an interarrival time of less than 10 milliseconds.
Around 80\% of all tasks have an interarrival time less than 100 milliseconds, with 95\% less than a second.

Overall, this demonstrates all domains require scheduling operations to happen at the sub-second level, with industry having the highest need for well-performing schedulers.

\begin{figure}[t]
	\centering
	\includegraphics[width=0.9\columnwidth]{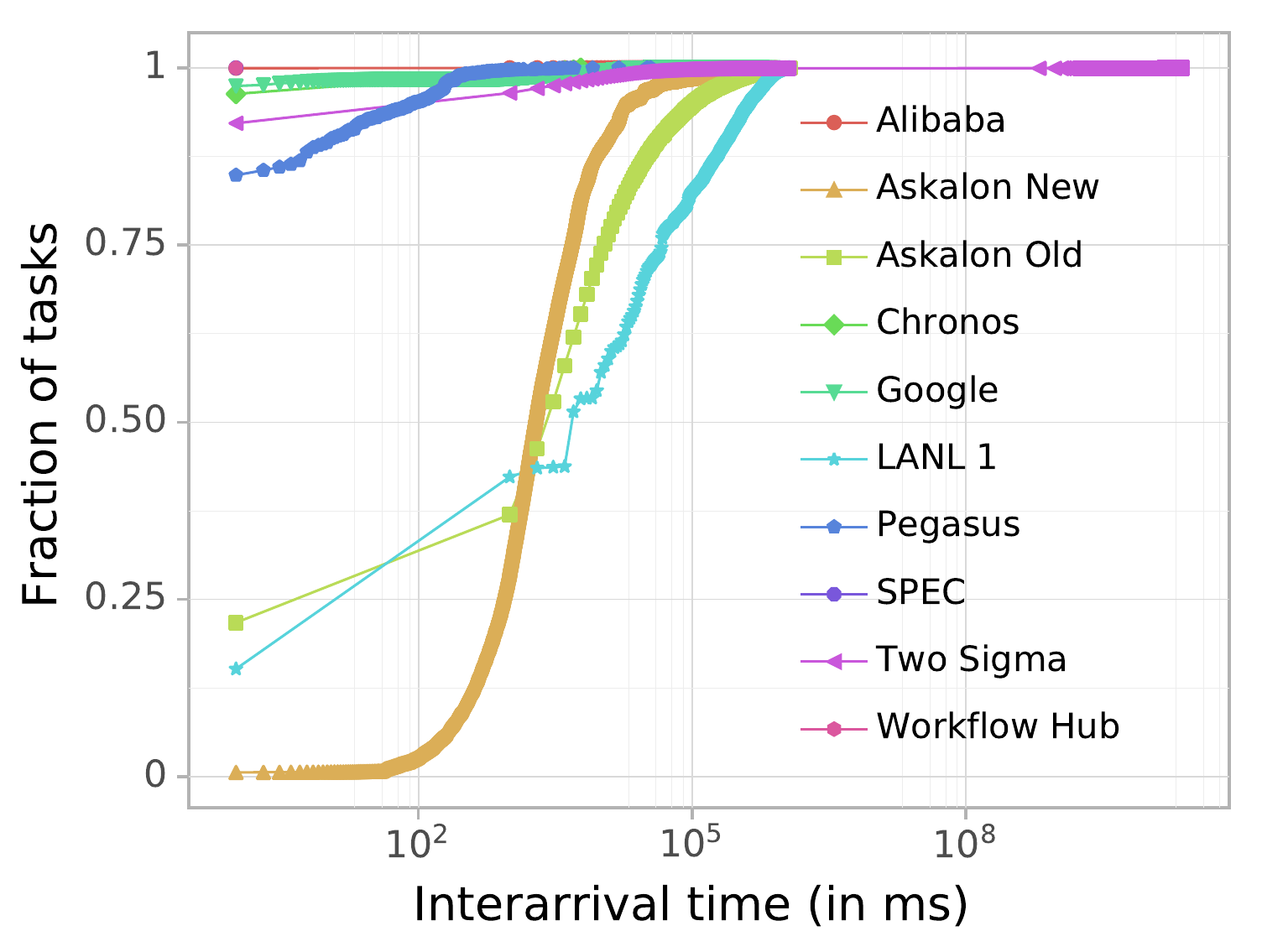}
	\vcutL
	\caption{CDF of task interarrival times, per source.} %
	\label{fig:exp_task-interarrival-time_source_trend}
	\vcutL
\end{figure}

To observe differences per use-case, Figure~\ref{fig:exp_task-interarrival-time_source_trend} depicts the task interarrival times per source.
From this figure we observe the Askalon (old and new) traces and LANL 1 exhibit mid-range interarrival times, as the majority of tasks have interarrival times between 10 milliseconds and 1 second.
All other sources show the majority of tasks have task interarrival times lower than 10 milliseconds, stressing the need for high-performance schedulers. 
While most tasks arrive quickly after one another, there are significant outliers in the Two Sigma traces.
This could indicate downtime of production systems, as the arrival pattern of Two Sigma is diurnal, yet stable (see Section~\ref{ssct:exp_longitudinal_trend}).

\section{Addressing Challenges of Validity}%
\label{sct:validation-and-threats}

In this section, we discuss challenges to the validity of this work. 
We address the challenges through either 
trace-based simulation (the first) or 
argumentation (the others).

{\bf Challenge} \mainchallenge{chal:trace-diversity}{Trace diversity does not impact the performance of workflow schedulers.}
As outlined in Sections~\ref{ssct:current-content} and \ref{sct:workflow-characterization}, the \gls{wta} traces are diverse.
However, {\it what is the impact of trace diversity?}

To demonstrate the impact of trace diversity on scheduler performance, we conduct a trace-based simulation study.
We simulate workloads from five sources using two scheduler configurations.
We equip the simulated scheduler with either the first-come first-serve (FCFS) or the shortest job first (SJF) queue sorting policy.
For both scheduler configurations, we further use a best-fit task placement policy.
We do not use a fixed resource environment to prevent bias when sampling or scaling traces~\cite{DBLP:conf/jsspp/FrachtenbergF05}.
Instead, we tailor the amount of available resources for each trace to reach roughly a 70\% resource utilization on average, based on the amount of CPU (core) seconds of trace and its length.
Although ambitious, 70\% resource utilization is achievable in parallel HPC environments~\cite{jones1999scheduling} and can be seen as a target for cloud environments.
To evaluate the performance of each scheduler, we use three metrics commonly used to assess schedulers' performance~\cite{DBLP:journals/jpdc/KwokA99,DBLP:conf/jsspp/FeitelsonRSSW97}: task response time (ReT), bounded task slowdown (BSD, using a lower bound of 1 second), and normalized workflow schedule length (NSL, the ratio between a workflow's response time and its critical path).

\begin{table}[t]
\caption{The performance in simulation of two schedulers for traces from different sources. Lower values are better.}\label{tbl:average-njsl}
\vspace*{-0.35cm}
\adjustbox{max width=\linewidth}{
\begin{tabular}{@{}llrrrrr@{}}
\toprule
 & & \multicolumn{5}{c}{Source of Trace} \\ \cmidrule{3-7}
Metric & Policy & Askalon Old & Askalon New & Pegasus & Shell & SPEC \\ \midrule
\multirow{2}{*}{Avg. ReT} & FCFS & $2.02 \cdot 10^5$ s & 167 s & $2.43 \cdot 10^4$ s & 9.76 s & 491 s \\
 & SJF & $1.74 \cdot 10^5$ s & 113 s & $2.12 \cdot 10^4$ s & 9.52 s & 248 s \\ \midrule
\multirow{2}{*}{Avg. BSD} & FCFS & $1.53 \cdot 10^4$ & 65.1 & $1.31 \cdot 10^3$ & 1.13 & 47.4 \\
 & SJF & $0.14 \cdot 10^4$ & 11.6 & $0.10 \cdot 10^3$ & 1.06 & 2.2 \\ \midrule
\multirow{2}{*}{Avg. NSL} & FCFS & $1.05 \cdot 10^5$ & 2.50 & $2.35 \cdot 10^3$ & 1.12 & 13.9 \\
 & SJF & $0.01 \cdot 10^5$ & 3.14 & $0.06 \cdot 10^3$ & 1.07 & 1.78 \\ \bottomrule
\end{tabular}
}
\vspace*{-0.45cm}
\end{table}

We report the performance of each simulated scheduler in Table~\ref{tbl:average-njsl} per source.
From this table we observe significant differences between schedulers and trace sources.
In particular, we find that the relative performance of schedulers differs between trace sources.
For example, SJF outperforms FCFS on the normalized schedule length metric by up to two orders of magnitude on traces from Askalon Old and Pegasus.
In contrast, on traces from Askalon New and Shell, the scheduling policies perform similarly.
For other metrics, these differences are present, but less pronounced.
SJF performs better than FCFS on response time and slowdown for each trace source, but the differences in performance between the schedulers vary greatly across traces.

Overall, 
we kept the
working environment
fixed per trace, yet obtained significantly different results depending on the scheduler and input trace.
Thus, 
our trace-based simulations give practical evidence that researchers require experimenting with different traces to claim generality and feasibility of their proposed approaches.

\mainchallenge{chal:venue-selection}{Limited venue selection in the survey.}
Besides omitting venues that yielded no results on our initial query, we made sure that journals, workshops, and conferences were covered at various levels in term of quality.
We believe this covers the field of systems community to a degree where conclusions can be drawn from.
We specifically focus on articles published in the systems communities as specialized communities, e.g., bioinformatics, focus on systems that solve domain-specific problems, but rarely conduct in-depth experiments, including trace-based, to test the system-level capabilities and behavior.

\mainchallenge{chal:data-anonymization}{Level of data anonymization.}
The Google team published interesting work data~\cite{clusterdata:Reiss2012}, but their anonymization approach, of normalizing values of both resource consumption and available resources, reduces significantly the usability of traces and the characterization details they provide.
We argue this type of anonymization is not preferred.
When available resources per machine, e.g. available disk space, memory, etc., and resource consumption numbers are normalized, reusing traces for different environments becomes difficult.
Researchers then need to make assumption on what kind of hardware the workflows were executed as done in the work of Amvrosiadis et al.~\cite{DBLP:conf/usenix/AmvrosiadisPGGB18} or need to assume a homogeneous environment. %

Instead, obfuscation techniques, such as multiplying both consumption and resources by a certain factor, allow for relative comparisons and the possibility to replay scheduling the workload on the resources while still concealing the original data.

\mainchallenge{chal:workflow-trace-format}{The Workflow Trace Format}.
A fourth challenge is the properties included in the workload trace format.
For each encountered property in other formats, we carefully decided whether to include it or not.
Low-level details such as page caches are omitted to not unnecessarily complicate the traces.
If future work demands change, the versioning schema per object will allow for these additions.

\section{Related Work}

We survey in this section the relevant body of work focusing on trace archives and on characterizing workloads.
Differently from other archives, the \gls{wta} focuses on {\it workloads of workflows}, preserving workflow-level arrival patterns and task inter-dependencies not found in other archives. 
Differently from other characterization work, ours is the first to reveal and compare workflow characteristics across different domains and fields of application. %

{\bf Open-access trace archives:}
Closest to this work is WorkflowHub~\cite{da2014community}, which archives traces of workflows executed with the Pegasus workflow engine %
and offers them in a unifying format containing structural information.
WorkflowHub also provides a tool to convert Pegasus execution logs to traces, similar to our parsing tools.
Different from this work, WorkflowHub's
traces include a single workflow and thus not a workload with a job-arrival pattern.  %
WorkFlowHub also does not provide statistical insights per trace and thus, they do not meet requirements \wtareq{1} and \wtareq{3}, and only partially meet \wtareq{4}.

Also relatively close to this work, the ATLAS repository maintained by the Carnegie Mellon University~\cite{amvrosiadis2018diversity} contains two traces (the S3 traces in this work), with other two traces announced but not yet released (as announced, the S7 traces in this work). 
None of their published traces contains task-interdependency data, so, although overlapping with our S3 and S7, the ATLAS work is different in scope and in particular does not address workflows.
Further, they do not consider different domains nor fields, and their archive lacks a unified format, statistical insights, selection mechanisms, and tooling---thus, they do not meet our requirements {\bf R1--4}.

Other trace-archives with similarities to this work include 
the MyExperiment archive~(ME)~\cite{goble2007myexperiment},
the Parallel Workloads\\ Archive~(PWA)~\cite{feitelson2007parallel}, and %
the Grid Workloads Archive~(GWA)~\cite{DBLP:journals/fgcs/IosupLJADWE08}. %
ME stores workflow executables, and semantic and provenance-data, but not provide execution traces as \gls{wta} does and thus has different scope.
The PWA includes traces collected from parallel production environments, which are largely dominated by tightly-coupled parallel jobs and, more recently, by bag-of-tasks applications. 
The GWA 
includes traces collected from grid environments; differently from this work, these traces are dominated by bag-of-tasks applications and by virtual-machine lease-release data.

{\bf Workload characterization, definition, and modeling:} There is much related and relevant work in this area, from which we compare only with the closely related; other characterization work does not focus on comparing traces by domain and does not cover a set of characteristics as diverse as this work, leading to so many findings.
Closest to this work, the Google cluster-traces have been analyzed from various points of view, e.g., \cite{reiss2012heterogeneity, chen2010analysis, mishra2010clustering}.
Amvrosiadis et al.~\cite{amvrosiadis2017bigger,amvrosiadis2018diversity} compare the Google cluster traces with three other cluster traces, of 0.3-3 times the size and 3-60 times the duration, and find key differences; our work adds new views and quantitative data on diversity, through both survey and characterization techniques.
Bharathi et al.~\cite{bharathi2008characterization} provide a characterization on workflow structures and the effect of workflow input sizes on said structures.
Five scientific workflows are used to explain in detail the compositions of their data and computational dependencies.
Using the characterization, a workflow generator generator for parameterized workflows is developed.
Juve et al.~\cite{juve2013characterizing} provide a characterization of six scientific workflows using workflow profiling tools that investigate resource consumption and computational characteristics of tasks.
The teams of Feitelson and Iosup have provided many  characterization and modeling studies for parallel~\cite{DBLP:journals/jpdc/FeitelsonTK14}, grid~\cite{DBLP:journals/internet/IosupE11}, and hosted-business~\cite{shen2015statistical} workloads; and Feitelson has written a seminal book on workload modeling~\cite{DBLP:books/daglib/0036899}.
In contrast, this work addresses in-depth the topic of workloads of workflows.

\section{Conclusion and Ongoing Work}\label{sec:conclusion}

Responding to the stringent need for diverse workflow traces, in this work we propose the \acrfull{wta}, which is an open-access archive containing workflow traces.

We conduct a survey of how the systems community uses workflow traces, by systematically inspecting articles accepted in the last decade in peer-reviewed conferences and journals.
We find that, from all articles that use traces, less than 40\% use realistic traces, and less than 15\% use any open-access trace. Additionally, the community focuses primarily on scientific workloads, possibly due to the scarcity of traces from other domains. These findings suggest existing limits to the relevance and reproducibility of workflow-based studies and designs.

We design and implement the \gls{wta} around five key requirements. At the core of the \gls{wta} is an unified trace format that, uniquely, supports both workflow- and task-level \glspl{nfr}. The archive contains a large and diverse set of traces, collected from 10 sources and encompassing over 48 million workflows and 2 billion CPU core hours.

Finally, we provide deep insight into the \gls{wta} traces, through a statistical characterization revealing that: 
(1) there are large differences in workflow structures between scientific, industrial, and engineering workflows, 
(2) our two biggest traces-- from Alibaba and Google-- have the most stable arrival patterns in terms of tasks per hour,
(3) industrial workflows tend to have the highest level of parallelism, 
(4) the level of parallelism per domain is clearly divided,
(5) engineering workloads tend to have the most tasks on the critical path, 
(6) the three domains inspected in this work show distinct critical path curves,
(7) in order to claim generality of an approach, one should test a system with a variety of traces with different properties, possibly from different domains. 

In ongoing work, we aim to attract more organizations to contribute real-world traces to the \gls{wta}, and to encourage the use of the \gls{wta} content and tools in educational and production settings.
One of our goals is to develop a library system administrators can integrate into their systems to generate traces in our format.
Our preliminary experience with this learns that developing such a library, even for a single system, requires significant engineering effort and is thus left for future work.
We aim to support other formalisms in the future, including directed graphs, BPMN workflows, etc. based on the community's needs.
Furthermore, we aim to improve the trace format and statistics we report for each trace, based on community feedback.

\section*{Reproducibility Statement}

The \gls{wta} datasets are available online on the archive's website \newline \url{https://wta.atlarge-research.com/}. The \gls{wta} tools. simulator, and parse scripts are available as free open-source software at \newline \url{https://github.com/atlarge-research/wta-tools},\newline \url{https://github.com/atlarge-research/wta-sim} and \newline \url{https://github.com/atlarge-research/wta-analysis}, respectively.

\bibliographystyle{plain}
\bibliography{bibliography_beautified.bib}

\appendix

\section{Artifact Evaluation and Description}
We present in this appendix the artifacts (data and tools) used in this work, where to find them, and how to inspect and rerun the tools on said data to reproduce and verify our findings.

\subsection{Traces}

All traces are available on the \gls{wta} archive website\footnote{URL removed to comply with SC19 double-blind standards.}.
The traces are available without restrictions and can be redistributed.

\subsubsection{Survey Workflow Trace Usage}

The data used in the survey and the tools used to visualize the data are available as open-source software at \textit{\url{https://github.com/atlarge-research/wta-analysis}}.
The data of survey can be found in the file \textit{\url{Literature_survey-usage_of_WFs-2009-2018_2019-05-23.csv}}.
All visualizations regarding this data were generated using the \textit{\url{parse_survey_csv.py}} Python script, which requires python 3, and the Python libraries NumPy (1.16.2) and Pandas (v0.24.2) to execute.

\subsubsection{Parsing traces}
To parse traces into parquet file format with the \gls{wta} workload format, the archive offers a parse script for each distinctive workload structure at \textit{\url{https://github.com//atlarge-research/wta-tools}}.
All parse scripts are available as open-access data and can be inspected for correctness and used as a foundation for other parse scripts.
All parse scripts are located in the \textit{\url{parse_scripts/parquet_parsers/}} folder and the filenames are indicative of which raw data source they parse. For example, the \textit{\url{workflowhub_to_parquet.py}} Python script parses a trace from the WorkflowHub archive.
All output can be found in \textit{\url{parse_scripts/output_parquet/}}, with a sub-folder for the source.
Thus, the \textit{\url{workflowhub_to_parquet.py}} Python script will output to\\ \textit{\url{parse_scripts/output_parquet/workflowhub/}}.
\subsubsection{Validating Traces}
To validate the traces and their properties, the archive offers a trace validation tool.
This tools can be found in the GitHub repository containing \gls{wta} tools.
The tool expects the traces in parquet file format using the unified trace format introduced in this work.
The run the tool, run the \textit{\url{validate_parquet_files.py}} Python script with as argument the path to the trace directory to be validated.
The tool can be found in the \textit{\url{parse_scripts}} directory.
The tool will print the results to standard output (or stdout) and will exit with a non-zero exit code.

\subsection{Trace Characterization}
All scripts used to characterize the contents of the \gls{wta} are available online at \textit{\url{https://github.com//atlarge-research/wta-analysis}}.
These scripts are offered as IPython notebook files, which includes both the code and the visual outputs, i.e. graphs.
In order to run all IPython notebook files, JuPyter (4.1) is required and the python libraries PySpark (2.4.3), plotnine (0.5.1), NumPy (1.16.2), Pandas (v0.24.2), more-itertools (7.0.0), and hurst (0.0.5) should be installed on the machine running JuPyter.

\subsubsection{Simulation Experiment}
The simulator used in Section~\ref{sct:validation-and-threats} is available online as open-source software at \textit{\url{https://github.com//atlarge-research/wta-sim}}.
The datasets used are available in the open-access archive introduced in this work.

The simulator is written in Kotlin 1.3 and is executed on a system running CentOS Linux release 7.4.1708 and an Intel(R) Xeon(R) CPU E5-2630 v3 @ 2.40GHz. The parquet data was read using the Avro parquet reader (1.10.1) from the Hadoop (3.2.0) project.

To rerun the simulation experiment, download the simulator sources from the location provided above, compile the simulator using Maven (\textit{mvn package}), and run the following command for each trace and simulator policy:

{
\begin{verbatim}
  java -cp target/wta-sim-0.1.jar \
    science.atlarge.wta.simulator.WTASim \
    -i "/path/to/traces/${trace_name}_parquet/" \
    -o "simulation_results/${trace_name}/${policy}" \
    --target-utilization 0.7 \
    --task-placement-policy best_fit \
    --task-order-policy ${policy}
\end{verbatim}
}

Replace the \textit{trace\_name} variable with the name of a trace to analyze (e.g., \textit{askalon\_ee}) and replace the \textit{policy} variable with \textit{fcfs} or \textit{sjf} for the appropriate queue sorting policy. The data used to create Table~\ref{tbl:average-njsl} can be found in \textit{\url{simulation_results/${trace_name}/${policy}/summary.tsv}}. Additional information about using the simulator is provided in the accompanying \textit{README.md} file.

\end{document}